\documentclass[submission, PhysLectNotes]{SciPost}

\usepackage{fontsize}
\changefontsize[12.4]{11}

\usepackage{graphicx,hyperref}
\usepackage{amsmath}
\usepackage{amsfonts}
\usepackage{epsfig} 
\usepackage{color}
\usepackage{bbm}
\usepackage{latexsym} 
\usepackage{amsmath}
\usepackage{amssymb}
\usepackage{amsthm}

\usepackage{tikz}
\usepackage{tikz-cd}
\usetikzlibrary{decorations.pathmorphing, shapes.misc}

\newcommand{\beq}{\begin{equation}}
\newcommand{\eeq}{\end{equation}}
\newcommand{\bea}{\begin{eqnarray}}
\newcommand{\eea}{\end{eqnarray}}

\mathchardef\nss="711B

\usepackage{braket}

\def\nss{G}

\def\be{\begin{eqnarray}}
\def\ee{\end{eqnarray}}

\newlength{\myL}

\renewcommand{\[}{\left[}

\def\scS{\mathcal{S}}

\def\Tr{\mathop{\mathrm{Tr}}}

\newcommand{\avg}{\mathop{\mathbb{E}}}

\newcommand{\eqnref}[1]{Eq.\,\eqref{#1}}
\newcommand{\figref}[1]{Fig.\,\ref{#1}}

\newcommand{\dia}[3]{\raisebox{#3pt}{\includegraphics[height=#2pt]{dia_#1}}}

\makeatletter
\newcommand\xleftrightarrow[2][]{%
  \ext@arrow 9999{\longleftrightarrowfill@}{#1}{#2}}
\newcommand\longleftrightarrowfill@{%
  \arrowfill@\leftarrow\relbar\rightarrow}
\makeatother

\hypersetup{
    colorlinks,
    linkcolor={red!50!black},
    citecolor={blue!50!black},
    urlcolor={blue!80!black}
}

\usepackage[bitstream-charter]{mathdesign}
\DeclareSymbolFont{usualmathcal}{OMS}{cmsy}{m}{n}
\DeclareSymbolFontAlphabet{\mathcal}{usualmathcal}

\begin{document}

\begin{center}{\Large \textbf{
Les Houches lectures on random quantum circuits and monitored quantum dynamics
}}\end{center}

\begin{center}
Romain Vasseur
\end{center}

\begin{center}
 Department of Theoretical Physics, University of Geneva, 24 quai Ernest-Ansermet, 1211 Gen\`eve, Switzerland
\\
Geneva Quantum Center, University of Geneva
\\
${}^\star$ {\small \sf romain.vasseur@unige.ch}
\end{center}

\begin{center}
\today
\end{center}


\section*{Abstract}
{\bf
These lecture notes are based on lectures given by the author at the  Les Houches 2025 summer school on ``Exact Solvability and Quantum Information''. The central theme of these notes is to apply the philosophy of statistical mechanics to study the dynamics of quantum information in ideal and monitored random quantum circuits --  for which an exact description of individual realizations is expected to be generically intractable. 
}

\vspace{10pt}
\noindent\rule{\textwidth}{1pt}
\tableofcontents\thispagestyle{fancy}
\noindent\rule{\textwidth}{1pt}
\vspace{10pt}

\section{Introduction}

Random quantum circuits have emerged as a powerful paradigm to explore the universal features of quantum dynamics in many-body systems. By combining unitary evolution drawn from random ensembles interspersed with local measurements, monitored circuits provide a versatile framework to study how quantum information can be encoded by quantum dynamics through scrambling, or how it can ``leak'' to the observer. In recent years, it was realized that this competition  between unitary entangling dynamics (that tends to scramble information) and projective measurements (that tend to reveal it) leads to measurement-induced phase transitions (MIPTs). These transitions are characterized by a change in the nature of quantum entanglement of quantum trajectories, or by the amount of information extracted by the observer. 

These Les Houches lecture notes are meant to provide a brief but self-contained introduction to entanglement dynamics in random quantum circuits and to measurement-induced phase transitions, with a focus on exact methods and mappings onto statistical mechanics models. 
 There are already two useful reviews on related topics~\cite{Potter2022, annurev:/content/journals/10.1146/annurev-conmatphys-031720-030658}, as well as some lecture notes from Sagar Vijay and myself based on lectures given at a Boulder summer school in 2023 (\href{https://boulderschool.yale.edu/2023/boulder-school-2023-lecture-notes}{link}), and from Brian Skinner~\cite{skinner2023lecturenotesintroductionrandom}, who gave a lecture series at the 2023  Condensed Matter Summer School at the University of Minnesota. The lecture notes below will inevitably overlap some of those resources (which I definitely encourage the interested reader to read for complementary perspectives on these topics). Complementary lectures on quantum circuits by Bruno Bertini were also given at the same Les Houches school~\cite{bertini2026nonequilibriumquantummanybodyphysics}: for this reason I will not discuss at all dual-unitary circuits. 
 
These lecture notes are not intended to be a review like Refs.~\cite{Potter2022, annurev:/content/journals/10.1146/annurev-conmatphys-031720-030658}. I will not attempt to give a comprehensive overview of all results in the field, nor will I attempt to provide an extensive review of the literature. In fact, I tried to keep the number of cited references to a minimum, to only provide key references directly relevant to results discussed in the notes. 
In order to fit a 3-lecture format, many topics are omitted, and these choices reflect the general theme of the summer school focusing on exact solutions as well as my own interests. I chose to discuss a select number of topics in detail, providing more pedagogical detail. However, Section~\ref{SecStatMech} will inevitably have some overlap with the review I wrote with Andrew Potter a few years ago~\cite{Potter2022}, with additional details and steps to make the derivation more pedagogical. Brian Skinner's lectures provide a beautiful introduction to measurement-induced phase transitions, with a focus on the so-called minimal cut picture and on analogies with percolation~\cite{skinner2023lecturenotesintroductionrandom}. In order to avoid redundancy, I tried to provide an orthogonal picture in terms of ``learnability'', and because of the school's theme of exact solutions, the notes below are focused on exact techniques and mappings onto replica statistical mechanics models.  Finally, the notes below will partly overlap with Sagar Vijay's Boulder lectures (especially in Sec.~\ref{SecRUC}), and my own Boulder lectures (although the Boulder lectures did not address learning aspects of measurement-induced phase transitions, and discussed extensively random tensor networks which I will not cover here).

I will assume some familiarity with concepts like entanglement entropy and tensor networks, which were discussed in other lectures at this Les Houches school. 

The outline of the lectures is as follows:
\begin{itemize}
\item In section~\ref{SecRUC} (lecture $\# 1$), we will first introduce the concept of unitary random quantum circuits focusing on entanglement growth under unitary dynamics~\cite{PhysRevX.7.031016,Nahum2018}. We'll discuss Haar averaging and illustrate how one can compute the average purity as a function of time in such circuits in terms of a mapping onto a classical Ising magnet~\cite{Nahum2018}. We will also discuss similarities and differences with random tensor network wavefunctions~\cite{PhysRevB.100.134203}. 

\item In section~\ref{SecMIPT} (lecture $\# 2$), we will switch gears and introduce measurement-induced phase transitions. Useful references include Refs.~\cite{PhysRevB.98.205136,Skinner2019,Potter2022, annurev:/content/journals/10.1146/annurev-conmatphys-031720-030658}. We will briefly present different perspectives on such transitions (as entanglement, purification, and learnability phase transitions), with an emphasis on how they quantify ``learning'' from measurement outcomes~\cite{Bao2020,PhysRevLett.129.200602}.
 
 \item Finally, in section~\ref{SecStatMech} (lecture $\# 3$), we will combine the results of the first two lectures and provide general statistical mechanics mappings to study the dynamics of quantum information in monitored quantum circuits~\cite{Bao2020,Jian2020}, using a ``replica trick''~\cite{Zhou2019,PhysRevB.100.134203} to deal with the nonlinearities and effective randomness in this problem. 
\end{itemize}

\section{Random quantum circuits and tensor networks}
\label{SecRUC}

Much of traditional many-body physics has focused on equilibrium phenomena: phases of matter, symmetry breaking, and criticality in systems at or near thermal equilibrium. In contrast, recent advances in atomic, molecular, and optical platforms, as well as superconducting and trapped-ion qubit arrays, now allow unprecedented control over isolated quantum systems far from equilibrium. These systems can be prepared in  low-entangled (typically product) initial states and evolved under programmable interactions, with high spatial and temporal resolution. A central theoretical challenge is therefore to identify universal features of non-equilibrium quantum dynamics. What aspects of quantum thermalization are robust to microscopic details? How does local information spread under unitary time evolution? Under what conditions does a closed quantum system approach thermal equilibrium, and how can this be diagnosed?

A key ingredient in addressing these questions is the notion of {\em scrambling}: under generic many-body unitary evolution, initially local quantum information becomes encoded in highly non-local degrees of freedom. Entanglement growth provides a sharp diagnostic of this process. Starting from a weakly entangled (e.g.\ product) state, generic dynamics typically leads to rapid linear growth of entanglement entropy, eventually saturating to a volume-law value. This growth not only signals thermalization, but also controls the computational complexity of simulating the system using tensor-network methods.

To isolate universal features of such dynamics, it is useful to consider minimally structured models that retain only the essential ingredients of generic quantum systems: {\em locality} and {\em unitarity}. This philosophy parallels that of random matrix theory, where complicated Hamiltonians are replaced by ensembles constrained only by symmetry. In the present context, this leads to the study of {\em random unitary circuits}, introduced in Refs.~\cite{PhysRevX.7.031016,Nahum2018}. These models provide analytically tractable yet remarkably faithful descriptions of chaotic quantum dynamics, while ignoring properties such as energy conservation.

\subsection{Brick-work random quantum circuits}

\begin{figure}[bt!] 
    \centering
    \includegraphics[width=0.5\textwidth]{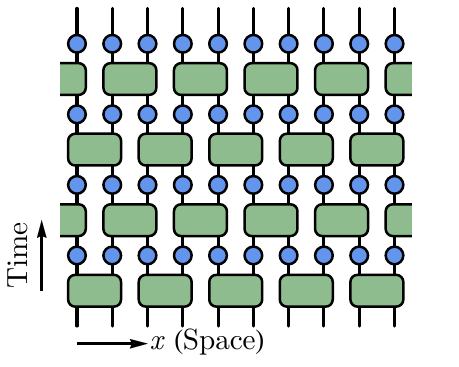}
    \caption{ {\bf Random quantum circuit. } Brick-work random quantum circuit evolution of a one dimensional system of qudits (horizontal direction), implementing a random discrete time evolution (vertical direction). Each green rectangle is a unitary gate acting on two nearest-neighbor qudits, drawn uniformly using the Haar measure. The blue dots represent potential measurements that will be discussed later in these notes.  Reproduced from Ref.~\cite{Jian2020}.
    \label{fig:circuit} 
            }
\end{figure}

We consider a one-dimensional system of $L$ qudits with Hilbert space ${\cal H}=({\mathbb{C}^{d}})^{\otimes L}$. The time evolution is given by a brickwork pattern of two-qudit gates acting on nearest-neighbor qudits, as shown in Fig.~\ref{fig:circuit}. The gates are drawn from the Haar (uniform) measure on $U(D=d^2)$. Recall that the Haar measure satisfies $\mathbb{E}_{U\in U(D)} f(U) = \mathbb{E}_{U\in U(D)} f(U V) = \mathbb{E}_{U\in U(D)} f(W U)$ for any unitaries $U,W \in U(D)$. We will be interested in typical properties of individual circuits. Importantly, note that even though the time evolution is random, the dynamics is  unitary. We will define a single time step $\Delta t=1$ as a layer of gates acting on even sites, followed by gates acting on odd sites.

You can think of these circuits as a Trotterized version of a random time-dependent Hamiltonian evolution, except that the gates need not be close to the identity. We will see that such circuits are extremely good ``toy models'' that are at the same time analytically tractable, but also generic enough to capture the universal features of chaotic (non-integrable) quantum dynamics with local interactions. This philosophy of constructing minimally-structured models (here using two key ingredients: locality and unitarity) should be familiar from random matrix theory. 

\subsection{Entanglement and minimal cut}

We will be interested in the entanglement dynamics in such random quantum circuits. Intuitively, starting from a product state, we expect entanglement to grow fast (linearly) with time, and to go to maximal (volume law) entanglement at long times. This is essentially thermalization at play: since there is no conserved quantity in the problem, reduced density matrices of a finite interval become maximally mixed (corresponding to ``infinite temperature'') under the dynamics. For a subregion $A$ of the system, we will characterize entanglement using R\'enyi entropies
\begin{equation}\label{eqRenyi}
S_A^{(n)} = \frac{1}{1-n} \ln {\rm tr} \rho_A^n,
\end{equation}
where $\rho_A = {\rm tr}_{\overline A} \rho$ is the reduced density matrix of the system in region $A$. The limit $n \to 1$ gives the von Neumann entanglement entropy, $S_A = - {\rm tr} \rho_A \ln \rho_A$. It is also useful to remember that $S_A^{(n+1)} \leq S_A^{(n)} \leq S_A^{(0)} = \ln {\rm rank} \rho_A  $.

Specific Rényi values are of particular interest. From a quantum information perspective, the most relevant value is the von Neumann entropy $S_A^{(1)}$, which quantifies the asymptotic rate at which Bell pairs can be distilled from, or used to create the state by local operations and classical communication.  It is the entropy most directly tied to quantum communication and error correction. Other indices emphasize different parts of the entanglement spectrum: $S_A^{(2)}=-\ln {\rm tr}\rho_A^2$ measures the purity and is particularly convenient for  calculations and experiments, $S_A^{(0)}$ counts the Schmidt rank, and $S_A^{(\infty)}=-\ln\lambda_{\max}(\rho_A)$ probes the largest Schmidt weight. In generic thermalizing systems these entropies often agree on the leading area-law versus volume-law classification, but their values and subleading scaling need not agree. In what follows, we will first focus on the limit $n=2$.

It will be convenient to represent $S_A^{(n)} $ in \eqnref{eqRenyi}  in terms of the expectation value of a permutation operator
$\scS_{n,A}$ acting on the $n$-fold replicated state as
\begin{equation}
S_A^{(n)} =\frac{1}{1-n}\ln\Tr\left((\ket{\psi}\bra{\psi})^{\otimes n}{\mathcal S}_{n,A}\right),
\end{equation}
where ${\mathcal S}_{n,A}$ depends on the choice of entanglement region $A$ and is defined as
\begin{equation}\label{eq:boundary def}
{\mathcal S}_{n,A}=\bigotimes_{x}\hat{g}_x,\quad
g_x=\left\{\begin{array}{ll}(12\cdots n), & x\in A,\\
{\rm identity} = e,
& x\in \bar{A}.\end{array}\right.
\end{equation}
Here $g_x$ labels the permutation on site $x$, and $\hat{g}_x=\sum_{\lbrace i \rbrace}\ket{i_{g_x(1)}i_{g_x(2)}\cdots i_{g_x(n)}}\bra{i_1i_2\cdots i_n}$ is its representation on the replicated on-site Hilbert space, {\it i.e.} on its $n$-fold tensor product. 
As indicated in the equation above, $g_x$ is the cyclic (identity) permutation when $x$ is in the region $A$ (when $x$ is in the region $\bar{A}$). In the following, we will also write operators as states in a doubled (operator) Hilbert space ${\cal H}_{\rm op} = {\cal H} \otimes {\cal H}^*$ with inner product $\braket{O_1 | O_2} = {\rm tr}  \ O_1^\dagger O_2$, and write 
\begin{equation} \label{eqRenyiPermutations}
S_A^{(n)} = \frac{1}{1-n}\ln  \braket{ {\mathcal S}_{n,A}| \rho^{\otimes n}}  = \frac{1}{1-n}\ln  \braket{ {\mathcal S}_{n,A}|(\mathcal{U}\otimes \mathcal{U}^*)^{\otimes n} | \rho_0^{\otimes n}}  ,
\end{equation}
with $\rho_0$ the initial density matrix, $\mathcal{U}$ the total time-evolution unitary matrix, $\ket{{\mathcal S}_{n,A}} = \otimes_{x=1}^L \ket{g_x}$  and  the (unnormalized) permutation states:
\begin{equation} \label{eqPermutationStates}
\ket{{g_x}}=\sum_{\lbrace i \rbrace}\ket{i_1 i_{g_x(1)} i_2 i_{g_x(2)}\cdots i_n i_{g_x(n)}} \in ({\mathcal H} \otimes {\mathcal H}^*)^{\otimes n},
\end{equation}
with ${\mathcal H} = \mathbb{C}^{d}$  the on-site Hilbert space.

\begin{figure}[bt!] 
    \centering
    \includegraphics[width=0.5\textwidth]{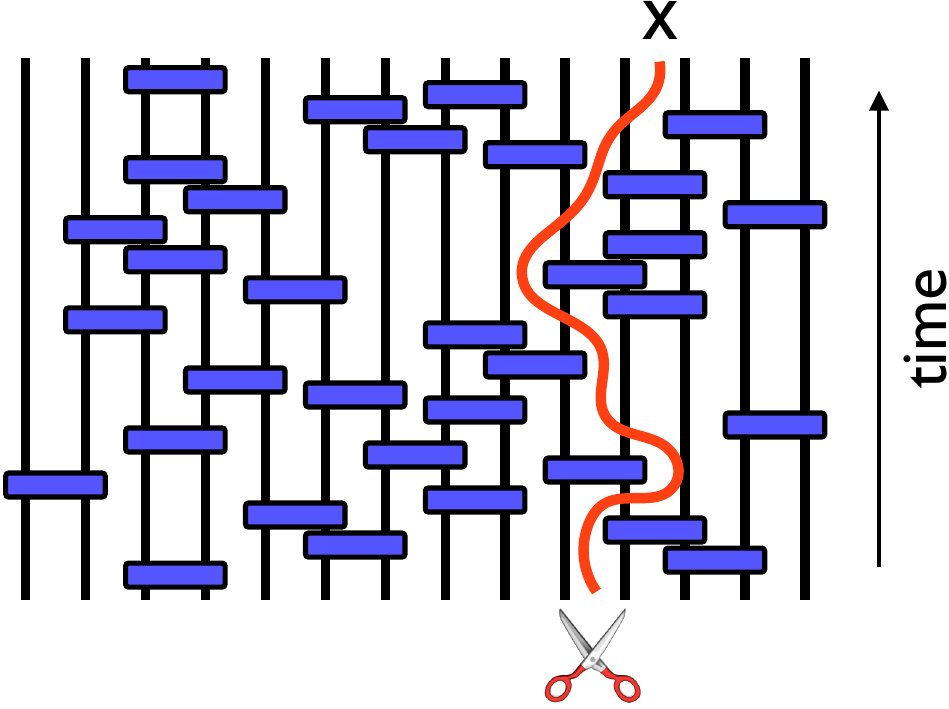}
    \caption{ {\bf Minimal cut through a quantum circuit.} A quantum circuit can be viewed as a tensor network representation of the state created by the circuit: each gate is a 4-legged tensor with a special structure that enforces unitarity. The quantum entanglement of a subregion $A$ is upper bounded by the number of links that need to be cut to isolate region $A$   times the logarithm of the bond dimension.   
     Reproduced from Ref.~\cite{PhysRevX.7.031016}.
    \label{fig:mincut} 
            }
\end{figure}

A useful tool to upper bound the entanglement entropy of a subregion $A$ at time $t$ is to interpret the circuit as a tensor network and use the ``minimal cut'' construction, following Ref.~\cite{PhysRevX.7.031016}. The notion of a minimal cut is quite general for arbitrary tensor-network wavefunctions. The basic idea is the following. Given a tensor-network representation of a wavefunction and a subsystem $A$, we can draw a cut through the network that separates the tensors associated with $A$ from those associated with its complement $\overline{A}$ (see Fig.~\ref{fig:mincut}). Cutting the internal legs along this surface exposes a set of bond indices. We can then rewrite the wavefunction as
\begin{equation}
\ket{\Psi} = \sum_{\mu} \ket{\Psi_A^\mu}\ket{\Psi_{\overline{A}}^\mu},
\end{equation}
where $\mu$ labels the collection of internal bond indices exposed by the cut. If each leg has bond dimension $\chi$, and if the cut intersects $k$ legs, then $\mu$ runs over at most $\chi^{k}$ values.

This resembles a Schmidt decomposition, except that the states $\ket{\Psi_A^\mu}$ and $\ket{\Psi_{\overline{A}}^\mu}$ are not orthonormal in general. Nevertheless, the number of terms in this expansion bounds the Schmidt rank of the state across the bipartition. Since the rank of the reduced density matrix cannot exceed $\chi^{k}$, we obtain
\begin{equation}
S_A^{(0)} \le k \ln \chi,
\end{equation}
and therefore for any R\'enyi entropy,
\begin{equation}
S_A^{(n)} \le S_A^{(0)} \le k \ln \chi.
\end{equation}
We can optimize this bound by minimizing $k$ over all possible cuts separating $A$ from $\overline{A}$. The resulting ``minimal cut'' (which need not be unique) provides the strongest upper bound on the entanglement entropy. In the context of random unitary circuits, the internal bond dimension is simply the local Hilbert-space dimension, $\chi = d$. To obtain a lower bound on entanglement growth, however, one must go beyond this purely geometric argument and use the randomness of the circuit, averaging over Haar-distributed unitary gates.

\subsection{Interlude: Haar averaging}\label{secHaarTwoReplicas}

If we attempt to average a quantity like the purity ${\rm tr} \rho_A^2$ over realizations of the random gates, we find that we need to average $(U \otimes U^*)^{\otimes 2}$ over the Haar measure, where $U \in U(D=d^2)$ is a two-qudit gate. The Haar average is given as a sum over permutations
\begin{equation}
\mathbb{E}_U \left[ U_{i_1 j_1} U_{i_2 j_2} {U}^*_{i_1' j_1'} U^*_{i_2' j_2'} \right] =
\sum_{\sigma, \tau \in S_2} \delta_{i_1 i_{\sigma(1)}'} \delta_{i_2 i_{\sigma(2)}'} 
\delta_{j_1 j_{\tau(1)}'} \delta_{j_2 j_{\tau(2)}'} \mathrm{Wg}_D(\sigma^{-1} \tau), \label{eqHaarAverage}
\end{equation}
where the coefficients are called Weingarten functions
\begin{equation} \label{eqW}
\mathrm{Wg}_D( \uparrow = e) = \frac{1}{D^2 - 1}, \qquad
\mathrm{Wg}_D( \downarrow = (12)) = -\frac{1}{D(D^2 - 1)},
\end{equation}
with $D=d^2$, and $e$ and $(12)$ the identity and swap permutations in $S_2 \simeq \mathbb{Z}_2$, respectively. More generally the average $\mathbb{E}_{U\in U(D)}\left((U\otimes U^*)^{\otimes Q}\right) $ can be expanded onto permutations of the replicas because such permutations commute with the action of the unitaries, and are the only terms surviving the Haar average. You can convince yourself that this is indeed the case using the invariance of the Haar measure:
\begin{equation}
\mathbb{E}_U \left[ (U \otimes U^*)^{\otimes Q} \right] = \mathbb{E}_U \left[  (V U W  \otimes V^* U^* W^*)^{\otimes Q} \right], 
\end{equation}
for any unitaries $V,W \in U(D)$. 
The commuting actions of the unitary and permutation groups on a tensor product Hilbert space is a mathematical statement known as {\em Schur-Weyl duality}. 

Let us make this structure more transparent. For a single replica, we introduce the operator space ${\cal H} \otimes {\cal H}^*$ with ${\cal H} = \mathbb{C}^D=\mathbb{C}^d \otimes \mathbb{C}^d$ acting on two qudits.
In the 2-fold replicated Hilbert space $({\cal H} \otimes {\cal H}^*)^{\otimes 2}$, we define the permutation states
\begin{equation} \label{Z2permutations}
\ket{\uparrow \uparrow } = \sum_{ij} \ket{i i jj}, \quad \ket{\downarrow \downarrow } = \sum_{ij} \ket{i jj i}.
\end{equation}
The up state corresponds to a pairing $(1 1^*) (2 2^*)$ of the replicas (identity pairing), while the down state corresponds to $(1 2^*) (2 1^*)$ (swap pairing): those are two permutation states for two replicas, as defined in eq.~\eqref{eqPermutationStates}. We use the notation $\ket{\sigma \sigma} = \ket{\sigma} \otimes \ket{\sigma}$ with $\sigma = \uparrow, \downarrow$ to make the two-leg structure transparent, but each leg is acted on by the same permutation.  
 Importantly, these states are neither normalized, nor are they orthogonal (check this! It might be convenient to use graphical tensor network notations):
\begin{equation} \label{eqOverlaps}
\braket{\uparrow \uparrow | \uparrow \uparrow} =  \braket{\downarrow \downarrow|\downarrow \downarrow} = D^2= d^4, \ \ \braket{\uparrow \uparrow | \downarrow \downarrow} = D=d^2.
\end{equation}
The operator $\mathbb{E}_U \left[ (U \otimes U^*)^{\otimes 2} \right] $ acting on this replicated Hilbert space is simply a projector on such permutation states
\begin{equation} \label{Haar2replicas}
\mathbb{E}_U \left[ (U \otimes U^*)^{\otimes 2} \right]  = \frac{1}{d^4 - 1} \left( \ket{\uparrow \uparrow}\bra{\uparrow \uparrow} + \ket{\downarrow \downarrow}\bra{\downarrow \downarrow} - \frac{1}{d^2} \left(  \ket{\uparrow \uparrow}\bra{\downarrow \downarrow} +  \ket{\downarrow \downarrow}\bra{\uparrow \uparrow}  \right) \right). 
\end{equation}
Note that here, the ``kets''  represent outgoing legs of the unitaries,  the ``bras'' represent incoming legs, whereas the states represent operators in the doubled Hilbert space. 
You can check that the coefficients are entirely fixed by the conditions 
\begin{equation}
\braket{ \sigma\sigma | \mathbb{E}_U \left[ (U \otimes U^*)^{\otimes 2} \right] | \sigma' \sigma'} = \braket{\sigma\sigma | \sigma'\sigma'},
\end{equation}
with $\sigma, \sigma'= \uparrow ,  \downarrow $, which follow from the fact that $\mathbb{E}_U \left[ (U \otimes U^*)^{\otimes 2} \right]$ commutes with the permutations $\sigma$: $\mathbb{E}_U \left[ (U \otimes U^*)^{\otimes 2} \right] \ket{\sigma \sigma} = \ket{\sigma \sigma}$. %

\subsection{Purity calculation and entanglement growth}
\label{SecPurity}

\begin{figure}[bt!] 
    \centering
    \includegraphics[width=0.7\textwidth]{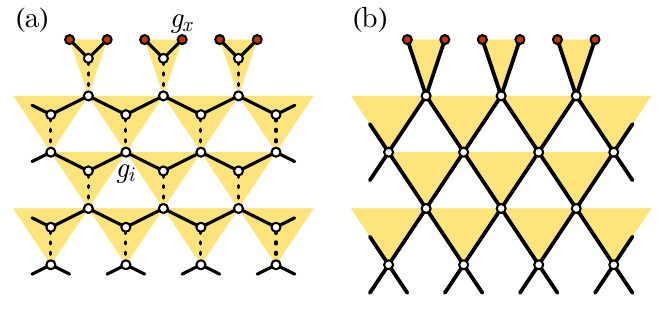}
    \caption{ {\bf Statistical mechanics model: geometry.} (a) Geometry of the statistical mechanics model of $S_Q$ spins. The red sites corresponds to the boundary spins to be pinned by the boundary condition. (b) In the $d=\infty$ limit, the model reduces to a Potts model on a square lattice. Reproduced from Ref.~\cite{Jian2020}.
    \label{fig:statmech} 
            }
\end{figure}

Following Ref.~\cite{Nahum2018}, let us apply this formalism to compute the average over unitaries of the purity at time $t$
\begin{equation}
Z_A  \equiv  \mathbb{E}_U {\rm tr} \rho^2_A(t),
\end{equation}
where $\rho_A(t) = {\rm tr}_{\overline{A}} \rho(t)$ with $ \rho(t) = \ket{\psi(t)} \bra{\psi(t)}$ the density matrix of the system starting from a pure state $\ket{0}$. Now, each two-qudit unitary gate can be replaced by~\eqref{Haar2replicas}. Each gate can be labelled by  two classical Ising spins $\sigma_i$ labelling the choice of permutation for incoming and outgoing legs. We find that $Z_A$ can be written as the partition function of an Ising model on an anisotropic hexagonal lattice shown in~\figref{fig:statmech}(a). The dashed links connecting sites vertically correspond to Weingarten weights~\eqref{eqW}, while the black links are weighted by the overlaps $\braket{\sigma_i | \sigma_j}$. Importantly, this Ising model has a global $\mathbb{Z}_2$ symmetry corresponding to flipping Ising spins: this invariance is already evident in eqs.~\eqref{eqOverlaps},\eqref{Haar2replicas}.

Let us introduce the  weight $J_p(\sigma_i,\sigma_j;\sigma_k)$ associated with
each down triangle (in yellow) in \figref{fig:statmech}(a),
\begin{equation} \label{eqTriangle}
J_p(\sigma_i,\sigma_j; \sigma_k)=\sum_{\sigma \in \lbrace \uparrow, \downarrow \rbrace } \braket{\sigma_i | \sigma}  \braket{\sigma_j | \sigma} \mathsf{Wg}_{d^2}(\sigma_k \sigma),
\end{equation}
which you can check happens to be positive. Those weights are given by
\begin{align}
J_p(\uparrow \uparrow; \downarrow)&= -\braket{\uparrow|\uparrow}^2 \frac{1}{d^2(d^4 - 1) } +  \braket{\uparrow|\downarrow}^2 \frac{1}{d^4 - 1 } =0, \notag \\
J_p(\uparrow \uparrow; \uparrow)&= \braket{\uparrow|\uparrow}^2 \frac{1}{d^4 - 1 } -  \braket{\uparrow|\downarrow}^2 \frac{1}{d^2(d^4 - 1) } =1,  \notag\\
J_p(\uparrow \downarrow; \uparrow)&= \braket{\uparrow|\uparrow} \braket{\downarrow|\uparrow} \frac{1}{d^4 - 1 } -  \braket{\uparrow|\downarrow} \braket{\downarrow|\downarrow} \frac{1}{d^2(d^4 - 1) } =\frac{d}{1+d^2},  \notag\\
\end{align}
where all other weights follow from symmetry.
The first line is a consequence of unitarity, and forces Ising domain walls to propagate upwards. 
The partition function can then be  written in terms of the triangle weight $J_p$ as
\begin{equation}\label{eq:ZA inf d}
Z_A =\sum_{\{ \sigma \}}\prod_{\langle ijk\rangle\in\triangledown}J_p(\sigma_i,\sigma_j;\sigma_k).
\end{equation}

\begin{figure}[bt!] 
    \centering
    \includegraphics[width=1.0\textwidth]{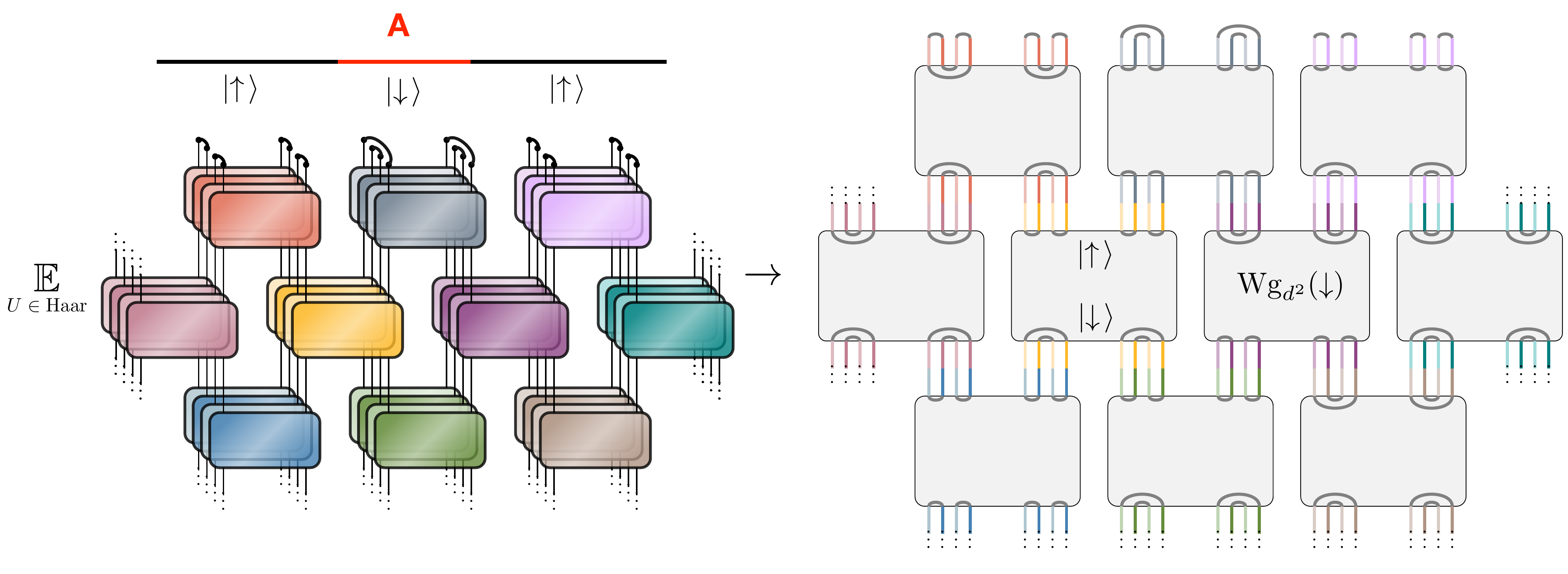}
    \caption{ {\bf Statistical mechanics mapping.} The calculation of the average purity in random unitary circuit dynamics 
    can be mapped onto an effective Ising model, where the two spins label different contractions of the two replicas. After averaging, the tensor network can be contracted exactly, giving local weights in terms of those Ising spins.
   Figure credits: Kabir Khanna, adapted from Ref.~\cite{khanna2025randomquantumcircuitstimereversal}.
    \label{fig:statmech2} 
            }
\end{figure}

What about the boundary conditions? Using the results of the previous section, we note that the purity can be written as
\begin{equation}
{\rm tr} \rho^2_A(t) =  \braket{ \sigma_A | \rho^{\otimes 2}},
\end{equation}
where $ \ket{ \rho^{\otimes 2}}$ is the state representing the replicated density matrix $\rho^{\otimes 2}$ in our replicated operator Hilbert space, and  $ \ket{ \sigma_A}$ is an Ising spin configuration with $\sigma_i = \uparrow$ if $i \in \overline{A}$ (identity permutation) and  $\sigma_i = \downarrow$ if $i \in A$ (implementing the swap permutation inside $A$). This fixes the boundary conditions at the ``top'' boundary (corresponding to final time $t$), see Fig.~\ref{fig:statmech2}. Our initial state satisfies $\braket{\uparrow | 0^{\otimes 4}}=\braket{\downarrow | 0^{\otimes 4}}=1$, so the ``bottom'' initial condition  (corresponding to the initial condition $t=0$) is free. 

As a sanity check, we can verify that 
\begin{equation}
Z_0 \equiv  \mathbb{E}_U ( {\rm tr} \rho(t) )^2 =1.
\end{equation}
The only difference in our Ising model is that in $Z_0$, the top boundary condition is fixed uniformly to $\sigma=\uparrow$. Since the weights forbid the creation of domain walls in the bulk, $Z_0$ is given by a single Ising configuration where all spins are up (our Ising model is deep in the ordered ferromagnetic phase). In contrast, the purity is given by the free energy cost of creating domain walls at the boundary of the interval $A$.  
\begin{equation}
\frac{Z_A}{Z_0} = {\rm e}^{- \Delta F_{\rm DW}},
\end{equation}
where $\Delta F_{\rm DW}$ scales as the length of the domain wall since it has a finite line tension (as the Ising model is ordered), so $\Delta F_{\rm DW} \sim t$ for $t \ll L_A$ and $\Delta F_{\rm DW} \sim L_A$ for $t \gg L_A$ for an entanglement interval of length $L_A$.  More explicitly, for $A$ corresponding to half of an infinite system, we have~\cite{Nahum2018}
\begin{equation} \label{eqPurityCircuits}
\mathbb{E}_U {\rm tr} \rho^2_A(t)  = \left(\frac{2d}{d^2+1} \right)^{2t} \sim {\rm e}^{- t / \tau_d},
\end{equation}
coming from a single domain wall starting from the top boundary to reach the bottom (Fig.~\ref{fig:unitary-domain-walls}). The domain wall can
choose between left-down or right-down moves as it proceeds from the top, and therefore there are $2^{2t}$ domain walls, each having exactly length $2t$.

We learn that the purity decays exponentially with time. Using the convexity of the function $- \ln x$, this result can be used to lower-bound the averaged 2nd R\'enyi entropy by $-\ln \mathbb{E}_U {\rm tr} \rho^2_A(t) $. Since the 2nd  R\'enyi entropy is also upper bounded by the minimal cut result $ 2 t \ln d$ (that is, $\ln d$  times the minimal number of cuts needed to disconnect the interval $A$ in the tensor network representing the system at time $t$), we conclude that the 2nd  R\'enyi entropy grows linearly with time. This ballistic growth of entanglement is a generic property of chaotic many-body quantum systems.  Analyzing the behavior of the averaged entanglement entropy requires a {\em replica trick}, which will be discussed in Sec.~\ref{SecStatMech}.

\begin{figure}[bt!]
    \centering
    \begin{tikzpicture}[x=0.48cm,y=0.38cm]
        \begin{scope}
            \clip (-6.2,-0.25) rectangle (6.2,6.25);
            \foreach \x in {-12,...,12} {
                \draw[black!18, line width=0.35pt] (\x,0) -- ({\x+6},6);
                \draw[black!18, line width=0.35pt] (\x,0) -- ({\x-6},6);
            }
        \end{scope}
        \draw[black, line width=0.7pt] (-6.2,6.25) -- (6.2,6.25);
        \draw[red!75!black, line width=1.4pt] (-0.55,6.25) -- (0.55,6.25);
        \node[above] at (-3.2,6.25) {$A$};
        \node[above] at (3.2,6.25) {$\overline A$};
        \node[above=2pt, red!75!black] at (0,6.25) {};
        \draw[purple!45, very thick]
            (0,6) -- (-1,5) -- (-2,4) -- (-1,3) -- (-2,2) -- (-3,1) -- (-2,0);
        \draw[purple!70, very thick]
            (0,6) -- (1,5) -- (2,4) -- (1,3) -- (2,2) -- (1,1) -- (2,0);
        \draw[purple!95!black, very thick]
            (0,6) -- (-1,5) -- (0,4) -- (1,3) -- (0,2) -- (1,1) -- (0,0);
        \fill[purple!75!black] (0,6) circle (2.2pt);
        \draw[<-, line width=0.7pt] (6.7,5.8) -- (6.7,0.4)
            node[midway,right] {$t$};
        \node[below] at (0,-0.1) {free initial boundary};
    \end{tikzpicture}
    \caption{{\bf Domain-wall paths contributing to Eq.~\eqref{eqPurityCircuits}.}
    Three representative paths are shown. The wall begins at the entanglement cut
    and, at each layer, independently takes a left-down or right-down step. A wall
    spanning $2t$ layers therefore has fixed length $2t$, while the sum contains
    $2^{2t}$ possible paths.}
    \label{fig:unitary-domain-walls}
\end{figure}

\subsection{Comparison with random tensor networks}

We see that under unitary dynamics, the entanglement entropy (more precisely we only considered the 2nd R\'enyi entropy) grows linearly with time, and approaches a volume-law scaling of entanglement at long times. What role did unitarity play in this result? Intuitively, non-unitarity might tame entanglement growth: for example, we already know that imaginary time evolution of the form ${\rm e}^{- \tau H}$ with $H$ a gapped Hamiltonian with unique groundstate will project onto the groundstate of $H$ at long times, which will be area-law entangled. 

How generic is this? We can get some insights into this question by replacing the unitary gates in the random quantum circuits by random {\em tensors} with bond dimension $d$ (that is, instead of matrix elements of random unitary gates, we now give independently sampled Gaussian random tensor elements). To compute a quantity like the purity, we have to face a major difficulty: the evolution does not preserve the normalization of the wave-function, so we have to normalize our state before computing anything physical. We could have avoided this by considering a more physical process like a random quantum channel, but averaging over random tensors is technically simpler.  
This requires a replica trick which we will explore in the next section in the context of measurement-induced phase transitions (see Ref.~\cite{PhysRevB.100.134203} for a discussion of statistical mechanics mappings using the replica trick in the context of random tensor networks). For simplicity, let us ignore this subtlety and assume that we can approximate the purity as

\begin{equation} \label{eqApproxRenyi}
 \mathbb{E}_{T} {\rm e}^{-S_2} = \mathbb{E}_{T}  \frac{   {\rm tr} \rho^2_A(t)}{({\rm tr} \rho(t))^2}    \approx \frac{  \mathbb{E}_{T} {\rm tr} \rho^2_A(t)}{\mathbb{E}_{T} ({\rm tr} \rho(t))^2} = \frac{Z_A}{Z_0},
\end{equation}
where $ \mathbb{E}_{T}$ denotes the average over random tensors (instead of random unitaries). If you are bothered (as you should be) by approximating the average of the ratio by the ratio of the averages in eq.~\eqref{eqApproxRenyi}, know that this turns out to be justified in the limit of large bond dimension~\cite{RTN}, and we will learn how to properly deal with the ratio later on in these lectures. How does the average over tensors compare to averaging over random unitaries? It turns out to be a lot simpler, we can just use Wick's theorem:
\begin{equation}
\mathbb{E}_T \left[ T_{\alpha_1} T_{\alpha_2} {T}^*_{\alpha'_1} T^*_{\alpha'_2} \right] =
\sum_{\sigma \in S_2} \delta_{\alpha_1, \alpha'_{\sigma(1)}}\delta_{\alpha_2, \alpha'_{\sigma(2)}},
\end{equation}
where $\alpha$ refers to the collection of indices (``legs'') of the tensors.  Comparing to~\eqref{eqHaarAverage}, we see that each tensor is associated with a single permutation (there's no distinction between incoming and outgoing legs here), which labels how kets and bras should be ``glued'', and there is no Weingarten function. Our statistical mechanics model is thus defined on a tilted square lattice (where each gate is replaced by a site of the square lattice), with partition function
\begin{equation}
Z = \sum_{\lbrace \sigma_i  = \uparrow, \downarrow \rbrace} \prod_{\langle i,j \rangle} \braket{\sigma_i | \sigma_j},
\end{equation}
with the boundary conditions in $Z_A, Z_0$ as before. The local Boltzmann weight leads to a ferromagnetic interaction that penalizes misaligned spins. The aligned overlaps $\braket{\uparrow | \uparrow }$ and $\braket{\downarrow | \downarrow }$ equal $d^2$, whereas the misaligned overlaps $\braket{\uparrow | \downarrow }$ and $\braket{\downarrow | \uparrow }$ equal $d$. This is just an Ising model again, now isotropic and defined on a square lattice, where $d$ controls the temperature:
\begin{equation}
{\cal H} = - \frac{ \ln d}{2} \sum_{\langle i,j \rangle} \sigma_i \sigma_j + {\rm const}.
\end{equation}
 We see that at large bond dimension $d\gg 1$ (low temperature), the statistical mechanics model is ordered, and we will again find volume-law entanglement: the key physics behind eq.~\eqref{eqPurityCircuits} is the non-zero line tension of the domain wall induced by the entanglement cut, which is a defining feature of ferromagnetic phases.
  At lower bond dimension, we have to use a replica trick, but our Ising model already gives us an important insight: lowering the bond dimension increases the effective temperature of the Ising magnet, so we can anticipate that eventually the statistical mechanics model will undergo a phase transition into a paramagnetic phase. In a paramagnet, the line tension of the domain walls (that lead to volume law scaling of entanglement) vanishes, so we expect an area-law phase. This transition in the statistical mechanics model corresponds to an {\em entanglement transition} in the random tensor network~\cite{PhysRevB.100.134203}. In the rest of these notes, we will see how those transitions can occur in random unitary circuits by introducing measurements.

\section{A brief introduction to measurement-induced phase transitions}
\label{SecMIPT}

Motivated by recent advances in noisy intermediate-scale quantum simulators~\cite{Preskill2018quantumcomputingin}, random quantum circuits have become a valuable tool to study how entanglement evolves in open quantum systems—systems that interact continuously with their environment. In this context, non-unitary random circuits naturally capture two competing effects: unitary evolution, which spreads and rapidly grows entanglement, and non-unitary processes such as measurements or environmental noise, which tend to destroy quantum information by revealing it~\cite{PhysRevB.98.205136,Skinner2019}. More generally, non-unitary circuits and random tensor networks~\cite{RTN,PhysRevB.100.134203} display different “phases” characterized by how entanglement scales with system size. A key example is the measurement-induced phase transition (MIPT)~\cite{PhysRevB.98.205136,Skinner2019}. This transition occurs in monitored random quantum circuits, which combine random unitary gates with local projective measurements applied at a fixed rate, separating two phases with distinct entanglement behavior.

To study MIPTs, we add local measurements into the mix, and study universal properties of monitored quantum dynamics. While it will be convenient to think of monitored phenomena in terms of quantum entanglement, we first focus on a more physically transparent ``learnability'' perspective in terms of how much information the observer is learning from the measurement outcomes.

\subsection{Learning from measurements: a simple example}

To illustrate the concept of learning from measuring a quantum state, we first turn to a simple example of distinguishability of two many-body states from a single-shot measurement in the computational basis. 
Let $\ket{\psi}$ and $\ket{\phi}$ be two random pure quantum states in a $D$-dimensional Hilbert space $(\mathbb{C}^d)^{\otimes L}$ with $D=d^L$ for a system of $L$ qudits. For $L$ large, these two states are essentially orthogonal. We assume both states are drawn randomly according to the Haar measure, and we are allowed to perform a \emph{single-shot measurement} in the computational basis $\{ \ket{m} \}$. We are given one of the two states, each with equal prior probability $1/2$, and we must guess which one it is based on the measurement outcome.

When a quantum state $\ket{\psi}$ is measured in the computational basis, the outcome $m$ occurs with probability $p(m|\psi) = \left| \braket{m|\psi} \right|^2$, and similarly for $\ket{\phi}$. Suppose we perform the measurement and obtain a given outcome $m$. We want to decide whether the state was $\ket{\psi}$ or $\ket{\phi}$. Using Bayes' theorem, the posterior probability that the state was $\ket{\psi}$ or $\ket{\phi}$ given outcome $m$ is:
\begin{equation} \label{eqBayesBorn}
P(\alpha=\psi,\phi | m) = \frac{p(m|\alpha)}{p(m|\psi) + p(m|\phi)},
\end{equation}
where we have used $p(\psi) = p(\phi) = 1/2$. We see that the best guess we can make is to select the state that has the higher likelihood for the observed outcome. That is, if the Born probabilities satisfy $p(m|\psi) > p(m|\phi)$, guess $\ket{\psi}$ since it was more likely to occur; otherwise, guess $\ket{\phi}$. This corresponds to a \emph{maximum-likelihood decision rule}, corresponding here to Bayes-optimal inference.

The probability of correctly guessing the state (averaged over all possible outcomes and both states) is given by:
\begin{align}
P_{\text{success}} &= \sum_{\lbrace m\rbrace} p(m|\psi)  \theta(p(m|\psi) - p(m|\phi)) p(\psi) +  p(m|\phi)  \theta(p(m|\phi) - p(m|\psi)) p(\phi), \notag \\ 
& =\frac{1}{2} \sum_{\lbrace m\rbrace}  \max\{ p(m|\psi), p(m|\phi) \}.
\end{align}
The first line has a direct physical interpretation: the true state is $\ket{\psi}$ with probability $1/2$, outcome $m$ then occurs with probability $p(m|\psi)$, and the step function is one precisely when the decision rule guesses $\psi$ correctly; and similarly for  the case if the true state is $\ket{\phi}$. The second term counts successful trials in which the true state is $\ket{\phi}$. Thus, for each $m$, only the larger of the two likelihoods contributes (ties have zero probability for random states and may be broken arbitrarily). Equivalently,
\begin{equation}
P_{\rm success}
=\frac{1}{2}+\frac{1}{2}D_{\rm TV}\!\left(p(\,\cdot\,|\psi),p(\,\cdot\,|\phi)\right),
\qquad
D_{\rm TV}(p,q)=\frac{1}{2}\sum_m |p_m-q_m|, \notag
\end{equation}
where the total variation distance is the standard statistical distinguishability
of two probability distributions.

 We now compute the ensemble mean of this quantity, which for this problem can easily be shown to coincide with its typical value (in the large $D$ limit). For random states, the Born probabilities $p(m|\psi) =  \left| \braket{m|\psi} \right|^2$ are distributed according to the so-called Porter-Thomas distribution
\begin{equation}
P[x= \left| \braket{m|\psi} \right|^2] \simeq D {\rm e}^{- D x},
\end{equation}
and similarly for $y=  \left| \braket{m|\phi} \right|^2$. Focusing on the large $D$ limit for simplicity, we see that the probability of success, or ``accuracy'', reads
\begin{equation} \label{eqMeanSuccess}
\mathbb{E}_{\psi,\phi}\!\left[P_{\text{success}}\right]
\underset{D \to \infty}{\sim}  \frac{D}{2}  \int_0^1 dx   \int_0^1 dy D^2 {\rm e}^{-D(x+y)}  \max\{ x, y \} = \frac{3}{4}.
\end{equation}
This might seem surprisingly high: there are exponentially ($d^L$) many states, each occurring with exponentially small Born probability,
and yet we're actually able to do a pretty good job at distinguishing the two states from the very limited information of the single round of measurements of the system. 

Equation~\eqref{eqMeanSuccess} is an ensemble statement about pairs of random states. A different question is how informative the posterior is from one experimental trial to the next. In each trial, let $\alpha_{\rm true}$ be the state actually prepared, sample $m$ from $p(m|\alpha_{\rm true})$, and define $p_{\rm corr}=P(\alpha_{\rm true}|m)$, the posterior probability assigned to the true label. Its probability density is
\begin{align}
P[p_{\rm corr}]  &= \sum_{\lbrace m\rbrace} p(m|\psi)  \delta \left( p_{\rm corr} - P(\psi|m)  \right) p(\psi) +  p(m|\phi)   \delta \left( p_{\rm corr} - P(\phi|m)  \right) p(\phi), \notag \\ 
& = D \int_0^1 dx   \int_0^1 dy D^2 {\rm e}^{-D(x+y)} x \delta\left(p_{\rm corr} - \frac{x}{x+y}\right) , \notag \\
& \simeq 2 p_{\rm corr}. 
\end{align}
The two terms on the first line are equal by exchange symmetry. The factor $x$ on the second line is a size bias: an outcome is sampled with the Born probability of the true state. We can check that the accuracy is $\mathbb{E}[p_{\rm corr}>\frac{1}{2}] = \frac{3}{4}$, while the  mean posterior weight placed on the true label (the ``credence'') is $\mathbb{E}[p_{\rm corr}] = \frac{2}{3}$.

\subsection{MIPTs as learnability transitions}

We now generalize this ``game'' of distinguishing initial states to many-body quantum dynamics. The unitary part of the dynamics is given by a random quantum circuit as in Sec.~\ref{SecRUC}, and after each layer of unitary gates, we measure each qudit in the computational basis with probability $p$. The probability $p$ controls the density of measurements in the dynamics: for $p=0$, there are no measurements, whereas for $p=1$, we measure the entire wavefunction after each layer of gates, projecting it onto a product state. 

We will call $\mathbf{m} = \lbrace 1,1,0,1,0,0, \dots \rbrace$ the measurement outcomes obtained from a given run of this monitored quantum evolution.  The corresponding state of the system $\rho_{\mathbf{m}} = \ket{\psi_{\mathbf{m}}} \bra{\psi_{\mathbf{m}}}$, labelled by the measurement outcomes, is called a quantum trajectory. The corresponding time evolution $\ket{\psi_{\mathbf{m}}} = K_{\mathbf{m}} \ket{0}$ is non-unitary, with the time-evolution operator $K_{\mathbf{m}} $ (called Kraus operators, satisfying $\sum_{\mathbf{m}} K_{\mathbf{m}}^\dagger K_{\mathbf{m}} =1$) comprising both unitary dynamics and projectors on measurement outcomes. We will use conventions where the state is not normalized after measurements, and the corresponding norm $p_{\mathbf{m}} = {\rm tr} \rho_{\mathbf{m}} $ is nothing but the Born probability to observe the outcomes $\mathbf{m}$ in the dynamics, with $\sum_\mathbf{m} p_{\mathbf{m}}  = 1$. 

\begin{figure}[bt!] 
    \centering
    \includegraphics[width=0.5\textwidth]{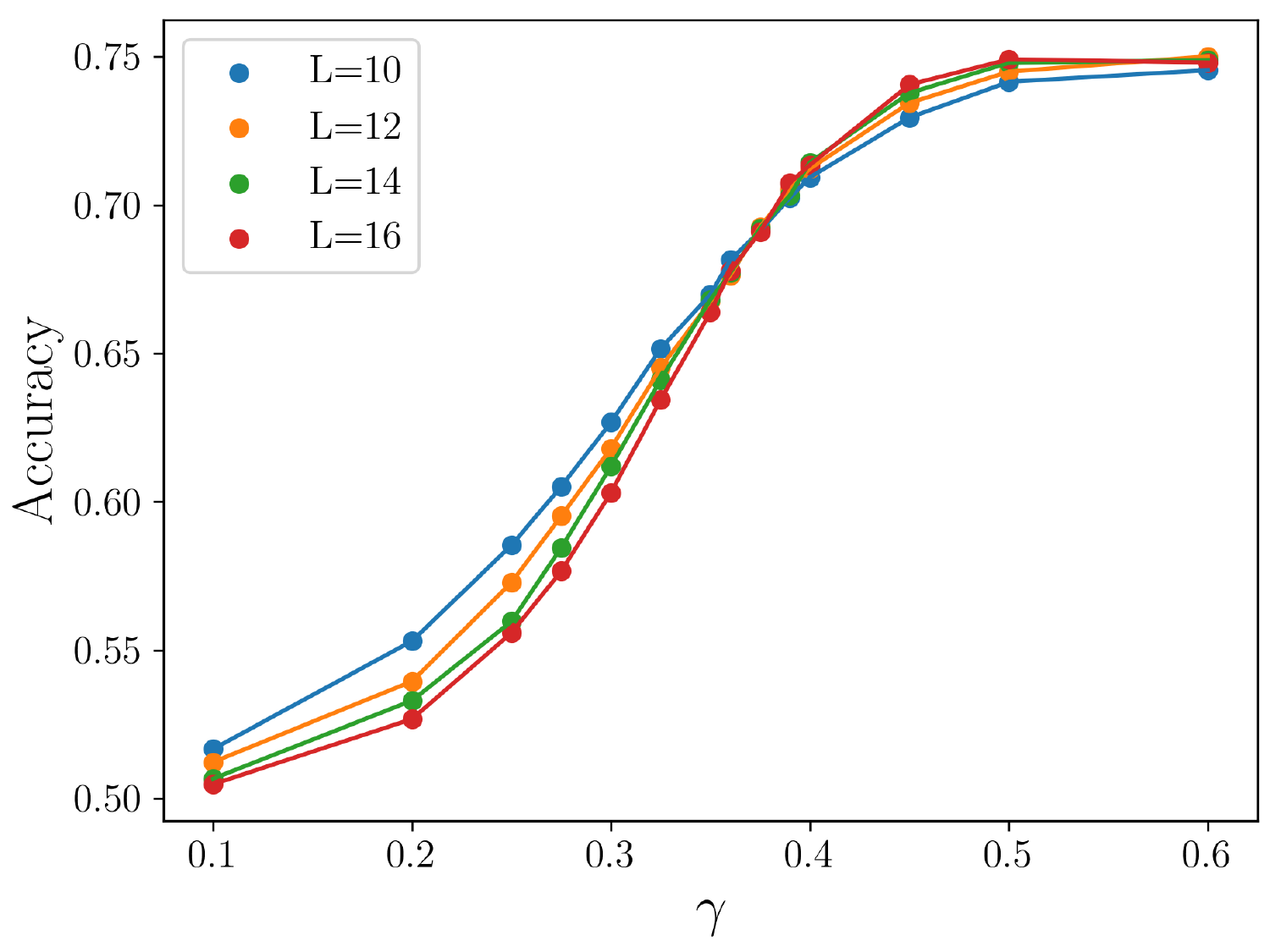}
    \caption{ {\bf Learnability transition. } Accuracy (probability of success) of distinguishing two random orthogonal initial states from the monitored dynamics (given by a random quantum circuit of depth $2L$) with measurement rate $\gamma$, for a system of $L$ qubits. (Figure credit: Abhishek Kumar, adapted from Ref.~\cite{kim2025learningmeasurementinducedphasetransitions}.) 
    \label{fig:accuracy} 
            }
\end{figure}

Importantly, the physics of MIPTs is not visible in the average density matrix $\overline{\rho}=\sum_\mathbf{m} \rho_{\mathbf{m}}$, but emerges in individual quantum trajectories—pure states conditioned on measurement outcomes—and in averages of nonlinear functions of the density matrix. One way to observe such transitions is to ask whether the observer can distinguish two orthogonal initial states~\cite{Gullans2020,PhysRevLett.129.200602} from the measurement outcomes $\mathbf{m}$ of the monitored random quantum circuits of depth of order $\sim {\rm poly}(L)$. In this setup, the observer knows everything about the initial states, the circuits and corresponding gates and measurement locations, and the measurement outcomes: their only task is to guess which initial state (chosen with prior probability $1/2$) was chosen. For this, we also assume their classical computational power is unlimited, so they can compute quantities such as Born probabilities $p_{\mathbf{m}}$ exactly (which in general is a hard task). 

For $p=1$, this reduces to the setup studied in the previous subsection: at time $t=0$, we measure the entire state and therefore destroy it, so the rest of the evolution does not reveal more information about the initial state. The probability to distinguish initial states is therefore $P_{\text{success}} = 3/4$ in that limit. On the other hand, for $p=0$, obviously the probability to distinguish two initial states is therefore $P_{\text{success}} = 1/2$ since the observer has no information to make their choice. The two endpoints alone do not imply a transition: in principle, the accuracy could vary smoothly between them. The physical reason to expect a finite threshold is
that both limiting behaviors are stable over a finite range of measurement rates. At low $p$, chaotic unitary dynamics scrambles the information about the two initial states into a
nonlocal encoding faster than dilute local measurements can read it: we will make this perspective more precise in the next section. For now, we will admit that  as a function of $p$, there is a phase transition occurring at a critical value of the measurement rate $p_c$. For $p<p_c$, the measurement outcomes contain no information in the thermodynamic limit, and we have $P_{\text{success}} = 1/2$. If the measurement rate is high enough $p>p_c$, the observer can start learning from the measurement outcomes and do better than flipping a coin to distinguish the two initial states. Numerical results obtained from exactly evaluating the Born probabilities in eq.~\eqref{eqBayesBorn} are shown in Fig.~\ref{fig:accuracy}. 
 
\subsection{Entanglement transitions}

MIPTs can also be probed in terms of entanglement of quantum trajectories -- this is how they were originally discussed in Refs.~\cite{PhysRevB.98.205136,Skinner2019}. In that language, MIPTs can be thought of as ``entanglement transitions'', which also occur in random tensor networks~\cite{PhysRevB.100.134203} (and aren't unique to measurement physics). 

When measurements are frequent enough, they efficiently extract quantum information, collapsing the system into weakly entangled states with area-law entanglement scaling (which you can loosely think of as a quantum Zeno effect). Conversely, at low measurement rates, unitary dynamics scramble information into nonlocal degrees of freedom that evade local measurements. This “entangling” or volume-law phase features highly entangled states starting from product states, and initially mixed states remain mixed for long times~\cite{Gullans2019}. This transition can be viewed either as a purification transition~\cite{Gullans2019} (see below) or through the lens of quantum communication and error correction: in the volume-law phase, the dynamics effectively protects information in a decoherence-free subspace, analogous to a quantum error-correcting code~\cite{Choi2020,Gullans2019,Li2020b,fan2020self}.

\subsection{Purification perspective} 

A product state evolving under monitored random quantum circuit dynamics exhibits one of two behaviors: either it develops extensive entanglement ($p < p_c$), or it retains only short-range entanglement due to measurement-induced collapse ($p > p_c$).

An insightful alternative approach is to consider feeding mixed states~\cite{Gullans2019} —or, equivalently, states initially entangled with external degrees of freedom—into monitored random quantum circuits. This viewpoint reveals rich connections between the entanglement phase transition and themes in quantum information, communication, and error correction.

To illustrate this, consider the ensemble of trajectories generated by inputting a maximally mixed (or ``infinite temperature'') state, $\rho_\infty = \frac{1}{d^L} \mathbbm{1}$, which carries an initial entropy of $S = -\text{tr}(\rho \ln \rho) = L \ln d$, into a monitored random circuit. In the extreme case where $p = 1$, every qubit is measured, and the system rapidly ``purifies'', collapsing the maximally mixed state into a pure state with zero  entropy. In contrast, for $p = 0$, the purely unitary evolution leaves the entropy unchanged, and the state remains maximally mixed at all times. Thus, mixed-state dynamics under monitored random circuits also undergo a phase transition: for $p > p_c$, frequent measurements purify any initial mixed state; for $p < p_c$, scrambling dynamics prevent purification, as it becomes ambiguous whether a measured qubit appears mixed due to entanglement with the environment or with other qubits in the system. This purification-based perspective closely relates to the ``learnability'' interpretation discussed earlier, though we will not elaborate on that connection here.

\begin{figure}[tb!] 
    \centering
    \includegraphics[width=0.85\textwidth]{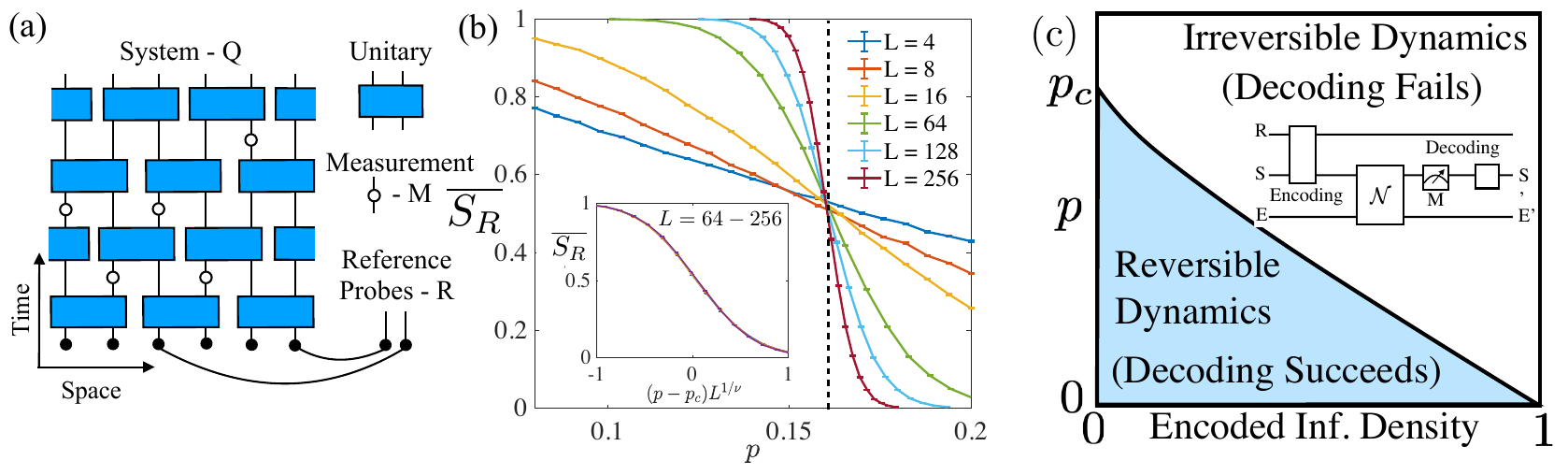}
    \caption{ {\bf Quantum information perspectives on the MIPT entanglement transition -- } adapted from \cite{Gullans2019}.
    (a) The entanglement transition can be interpreted as a purification transition for reference ancilla qubits (R) initially entangled with the system. The disorder-averaged entanglement entropy of the reference, $\overline{S_R}$, acts as an order parameter. When a pre-scrambling unitary is applied prior to the monitored dynamics, $\overline{S_R}$ exhibits a discontinuous change from $\ln(2)$ to $0$ in the thermodynamic limit, as supported by the finite-size scaling analysis shown in (b) for random Clifford circuits. (c) An equivalent viewpoint is provided by quantum communication theory: the pre-scrambling unitary functions as a random encoding of quantum information shared between $R$ and the system $S$, while the monitored dynamics $\mathcal{N}$ defines a noisy quantum channel, including monitoring by an environment $E$. One then considers decoding the information at the output using an optimal decoder. In this framework, the entanglement transition corresponds to a phase transition in the so-called ``quantum channel capacity'' of $\mathcal{N}$.
    \label{fig:ancilla} 
            }
\end{figure}

The maximally mixed input state $\rho_\infty$ can be understood as arising from a situation where each qubit in the system is entangled with an external ancilla qubit that remains isolated from the circuit dynamics. This purification perspective provides a powerful tool for characterizing the entanglement transition: one can track whether mutual information between the ancilla system and the evolving circuit persists at late times or is destroyed by measurement-induced collapse (see Fig.~\ref{fig:ancilla}). In fact, to detect the transition, it is sufficient to consider a single reference ancilla $R$ that is entangled with two orthogonal initial system states 
\begin{equation} \label{eqAncilla}
\ket{\Psi(t=0)} = \left(\ket{0} \ket{\phi} + \ket{1} \ket{\psi} \right)/\sqrt{2}.
\end{equation}
By examining the average trajectory of the reference ancilla across circuit realizations, one can extract a useful signature of the transition~\cite{Gullans2020}. Specifically, the entanglement entropy $\overline{S_R}$ of the reference ancilla exhibits a sharp jump across the transition point. This discontinuity serves as a practical numerical indicator: plotting $\overline{S_R}$ as a function of measurement rate $p$ for different system sizes $L$ reveals a crossing point, which precisely locates the entanglement transition.

The learnability, purification, and entanglement diagnoses are expected to locate the same transition because they track the same flow of initially encoded quantum
information. When the measurement record cannot distinguish the inputs $\ket{\phi} $ and  $\ket{\psi}$ in eq.~\eqref{eqAncilla}, the information remains coherently encoded in the
system: $R$ fails to purify, and the same protected many-body code space supports volume-law trajectory entanglement. When the record becomes informative, that
logical information has leaked to the observer; $R$ decouples and purifies, while measurements cut the code down to area-law trajectory states. Thus the three
observables differ at finite size but diagnose the loss of the same channel capacity in the thermodynamic limit~\cite{Gullans2019,Gullans2020,PhysRevLett.129.200602}.

\subsection{The ``post-selection problem'' and why it is not (really) a problem}
If you have heard about measurement-induced phase transitions (MIPTs), you may also have encountered what is sometimes called the ``post-selection problem.'' This terminology refers to the fact that MIPTs are only visible in quantities that are nonlinear in the density matrix $\rho_{\bf m}$ associated with a given measurement record ${\bf m}$. In other words, they are properties of individual quantum trajectories, and are not visible in the averaged density matrix
\begin{equation}
\overline{\rho} = \sum_{\bf m} p_{\bf m} \frac{\rho_{\bf m}}{{\rm tr} \rho_{\bf m}} = \sum_{\bf m}  \rho_{\bf m},
\end{equation}
which describes the system when measurement outcomes are ignored.

To understand why this matters, recall that in a realistic open quantum system, the environment effectively performs measurements on the system. However, unless we explicitly record the measurement outcomes, this information is lost to the environment. As a result, the state of the system is described by the averaged density matrix $\overline{\rho}$, and all physical observables must be computed from it. These observables are linear in the density matrix, for example
\begin{equation}
 \sum_{\bf m} p_{\bf m} \langle O \rangle_{\bf m}  = \sum_{\bf m} p_{\bf m} \frac{\langle  \psi_{\bf m} | O | \psi_{\bf m} \rangle}{\langle  \psi_{\bf m}  | \psi_{\bf m} \rangle}
= {\rm Tr}\left( \overline{\rho}\, O \right).
\end{equation}
Such linear quantities do not exhibit measurement-induced phase transitions.

By contrast, MIPTs appear in quantities that are nonlinear in the density matrix, such as
\begin{equation}
\sum_{\bf m} p_{\bf m} \langle O \rangle_{\bf m} ^2,
\end{equation}
or in entanglement entropies, which depend nonlinearly on $\rho_{\bf m}$. These quantities probe the properties of individual quantum trajectories before averaging over measurement outcomes.

The reason this is sometimes viewed as a ``post-selection problem'' is the following. To experimentally measure a nonlinear quantity such as the one above, one must repeatedly prepare the same quantum trajectory $\ket{\psi_{\bf m}}$, which requires obtaining the same measurement record ${\bf m}$ many times. Since measurement outcomes are random, the probability of observing any given trajectory decreases exponentially with the number of measurements. In our setup, the total number of measurements scales as $\sim p t L$, so post-selecting a specific trajectory becomes exponentially difficult in system size and time.

While this may sound like a serious limitation, it is more useful to think of it as a fundamental feature rather than a flaw. MIPTs belong to a broader class of information-theoretic phase transitions, which characterize how information flows, spreads, and can be recovered in many-body systems. Other examples include decoding transitions in quantum error correction and complexity transitions in random circuits and random tensor networks.

A useful analogy is provided by the toric code in the presence of noise. There, the central question is whether the encoded quantum information is preserved in the presence of noise, and how it can be recovered using a suitable classical decoding algorithm. This question defines a sharp phase transition between a phase where recovery is possible and a phase where it is not. Importantly, this transition is not defined in terms of directly measurable observables, but in terms of the information contained in the system and our ability to extract it.

Measurement-induced phase transitions can be understood in exactly the same spirit. They characterize how much information about the quantum state can be learned from the measurement outcomes. At low measurement rates, the measurement record contains too little information to learn anything about the quantum state: an observer would need an exponentially long time to even distinguish two different states. At high measurement rates, however, the measurement record contains sufficient information to infer some features of the state. 

From this perspective, the key question is not whether MIPTs can be observed directly in simple physical observables, but rather whether the measurement outcomes contain recoverable information about the quantum state. Answering this question requires classical post-processing (decoding) of the measurement record, rather than post-selection. This learnability or decoding perspective provides a natural and physically meaningful interpretation of measurement-induced phase transitions.

\section{Statistical mechanics mapping for monitored circuits}

\label{SecStatMech}

To gain a deeper and more systematic understanding of the physics of MIPTs, we now introduce an analytical approach that maps the dynamics of entanglement in these monitored quantum circuits onto a classical statistical mechanics problem, similar to what we did in Sec.~\ref{SecPurity} for the purity in unitary circuits. This mapping is made possible through a technique known as the replica trick, which is commonly used in disordered systems and statistical field theory.

We will focus on the ``entanglement transition'' perspective, where the MIPT separates phases with volume- and area-law scaling of the entanglement entropy. The key idea is that the entanglement entropy—particularly the averaged quantity over different circuit realizations—can be interpreted as a kind of free energy cost associated with creating a domain wall at the boundary of a classical spin system~\cite{PhysRevB.100.134203,Zhou2019,Bao2020,Jian2020}, like we already anticipated in Sec.~\ref{SecPurity}. Within this framework, the transition from a volume-law entangled phase to an area-law phase corresponds to a symmetry-breaking phase transition in the classical model. Specifically, it involves a breaking of replica symmetry, although of a different kind than the familiar type encountered in the study of spin glasses and disordered systems.

\subsection{Replica trick}

Our objective is to compute the \emph{Rényi entropies} associated with individual quantum trajectories, averaged over both measurement outcomes and realizations of random unitary gates. Each trajectory is labeled by a sequence of measurement outcomes $\mathbf{m}$ and is assigned a statistical weight given by its \emph{Born probability},
\begin{equation}
p_\mathbf{m} = {\rm tr}(\rho_{\mathbf{m}}),
\end{equation}
where
\begin{equation}
\rho_{\mathbf{m}} = \ket{\psi_{\mathbf{m}}}\bra{\psi_{\mathbf{m}}},
\end{equation}
is the (unnormalized!) pure-state density matrix associated with the trajectory $\mathbf{m}$. The reduced density matrix in a subsystem $A$ is defined as
\begin{equation}
\rho_{A,\mathbf{m}} = {\rm tr}_{\bar A} \rho_{\mathbf{m}} .
\end{equation}
Since we work with unnormalized density matrices during the time evolution, it is important to keep explicit normalization factors when computing physical quantities.

The $n$-th Rényi entropy of subsystem $A$, averaged over trajectories and random unitaries, is defined as
\begin{equation}
\label{eqDefRenyi}
S_{A}^{(n)}
=
\mathbb{E}_{U}
\sum_{\mathbf{m}}
p_\mathbf{m}
\frac{1}{1-n}
\ln\!\left[
\frac{{\rm tr}\!\left(\rho_{A,\mathbf{m}}^n\right)}
{\left({\rm tr}\rho_{\mathbf{m}}\right)^n}
\right],
\end{equation}
where $\mathbb{E}_{U}$ denotes the Haar average over random unitary gates. The sum over $\mathbf{m}$ runs over all quantum trajectories, i.e., all possible sequences of measurement outcomes. We will denote by $\overline{S_{A}^{(n)}}$ the Rényi entropy further averaged over the spatial positions of the measurements.

At first sight, evaluating Eq.~\eqref{eqDefRenyi} appears formidable. Entanglement entropies are already difficult to compute in interacting many-body systems, and here the problem is compounded by two additional ingredients: (i) the intrinsically non-equilibrium dynamics generated by random unitary evolution, and (ii) the nonlinearity introduced by measurements and Born-rule weighting. Following Refs.~\cite{PhysRevB.100.134203,Zhou2019,Bao2020,Jian2020}, we overcome this difficulty using the \emph{replica trick}, a method widely employed in the study of disordered systems and classical random spin models. The key identity underlying this approach is
\begin{equation}
\label{eqReplicaTrick}
\ln x
=
\underset{k \to 0}{\rm lim}\;
\frac{d}{dk}\,
x^k .
\end{equation}
This identity is exact. The strategy is to replace the logarithm—which is difficult to average—by powers $x^k$, compute the disorder average for integer $k$, and then analytically continue the result to $k \to 0$. The analytic continuation step can be subtle. For example, the function
\begin{equation}
f(n) = \frac{\sin(\pi n)}{\pi n},
\end{equation}
vanishes for all integers $n=1,2,\dots$, while
\begin{equation}
\lim_{n \to 0} f(n) = 1 .
\end{equation}
Thus knowledge of a function at integer values does not always uniquely determine its analytic continuation. Despite this caveat, the replica method has proven extremely powerful in the analysis of disordered and stochastic systems.

Applying Eq.~\eqref{eqReplicaTrick} to the Rényi entropy in Eq.~\eqref{eqDefRenyi}, we obtain
\begin{equation}
\label{eqSswap}
S_{A}^{(n)}
=
\underset{k \to 0}{\rm lim}\;
\frac{d}{dk}\,
\mathbb{E}_{U}
\sum_{\mathbf{m}}
\frac{p_{\mathbf{m}}}{1-n}
\left[
\left({\rm tr}\rho_{A,\mathbf{m}}^n\right)^k
-
\left({\rm tr}\rho_{\mathbf{m}}^{\otimes nk}\right)
\right].
\end{equation}
In this expression, the second term arises from the normalization factor $\left({\rm tr}\rho_{\mathbf{m}}\right)^n$ in Eq.~\eqref{eqDefRenyi}. It is convenient to introduce
\begin{equation}
Q = nk + 1 ,
\end{equation}
which represents the total number of replicas. The additional replica (the ``$+1$'') originates from the Born weight $p_{\mathbf{m}} = {\rm tr}(\rho_{\mathbf{m}})$ used to average over trajectories. Within this replicated formalism—where operators are represented as states in a doubled Hilbert space, as discussed in Sec.~\ref{SecRUC}—the entropy can be rewritten as
\begin{equation} \label{eqEntReplicaTemp}
S_{A}^{(n)}
=
\frac{1}{1-n}
\underset{k \to 0}{\rm lim}\;
\frac{d}{dk}\,
\mathbb{E}_{U}
\sum_{\mathbf{m}}
\left(
\braket{{\cal S}_{A,n} \mid \rho_{\mathbf{m}}^{\otimes Q}}
-
\braket{e \mid \rho_{\mathbf{m}}^{\otimes Q}}
\right).
\end{equation}
Here ${\cal S}_{A,n}$ is a permutation operator (a ``swap'' operator for $n=2$) acting on the replicated Hilbert space. It implements the cyclic permutation required to compute ${\rm tr}\rho_{A,\mathbf{m}}^n$ in the first $nk$ replicas, and acts as the identity on the additional replica associated with the Born weight.

More explicitly, we write
\begin{equation}
\label{eq:boundary def}
{\cal S}_{A,n}
=
\bigotimes_{x}
\ket{g_x},
\qquad
g_x=
\begin{cases}
(12\cdots n)^{\otimes k}, & x \in A, \\
e, & x \in \bar A .
\end{cases}
\end{equation}
Here $g_x$ specifies the permutation acting at site $x$, and $\ket{g_x}$ denotes the corresponding vector in the replicated on-site operator Hilbert space (now consisting of $Q = nk+1$ copies), see eq.~\eqref{eqPermutationStates}. For sites $x \in A$, the operator performs a cyclic permutation among the $n$ Rényi replicas within each of the $k$ groups; for $x \in \bar A$, it acts trivially. Throughout these lecture notes, we use standard cycle notation for permutations: for example, $(123)4$ denotes the permutation mapping $1234 \to 2314$. The essential point is that entanglement entropies are encoded as expectation values of permutation operators in a replicated theory, thereby transforming a nonlinear problem into a linear one in an enlarged Hilbert space.

\subsection{Haar calculus and Boltzmann weights}

The next step is to perform the average over the replicated random unitary gates using the Haar measure. We already encountered this procedure in Sec.~\ref{secHaarTwoReplicas} for the special case of two replicas; here we generalize the construction to an arbitrary number of replicas $Q = nk+1$. Haar averages over the unitary group play a central role in the theory of random quantum circuits, as they encode the statistical properties of fully random unitary evolution. For a pedagogical introduction and useful technical background, see Ref.~\cite{roberts2017chaos}.

The average over a single two-site unitary gate $U \in U(D)$, acting on a Hilbert space of dimension $D=d^2$, can be evaluated exactly using the Weingarten calculus. The relevant identity is
\begin{equation} \label{eqHaaraverage}
\mathbb{E}_{U\in U(D)}
\left(U\otimes U^{*}\right)^{\otimes Q}
=
\sum_{g_1,g_2 \in S_{Q}}
\text{Wg}_{D}(g_1^{-1}g_2)
\ket{g_1 g_1}
\bra{g_2 g_2},
\end{equation}
where $S_Q$ denotes the permutation group of $Q$ elements. The permutations $g_1,g_2$ act on the replica indices, and $\text{Wg}_{D}(g)$ denotes the Weingarten function associated with permutation $g$. The states $\ket{g_1 g_1} = \ket{g_1}\otimes\ket{g_1}$ represent permutation operators acting in the replicated operator Hilbert space (see eq.~\eqref{eqPermutationStates}). Concretely, $\ket{g_1 g_1}$ permutes the output legs of the unitary $U$ according to permutation $g_1$, and contracts them with the corresponding legs of $U^*$, while $\bra{g_2 g_2}$ similarly permutes and contracts the input legs. This structure reflects the fact that Haar averaging enforces permutation symmetry between replicas, and results in an effective interaction between replica degrees of freedom.

Using standard tensor network notation~\cite{ORUS2014117}, Eq.~\eqref{eqHaaraverage} can be represented graphically as~\cite{Jian2020}
\begin{equation} \label{eqHaaraverage2}
\avg_U \dia{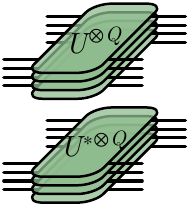}{56}{-26}
=
\sum_{g_1,g_2\in S_Q}
\text{Wg}_{D}(g_1^{-1}g_2)
\dia{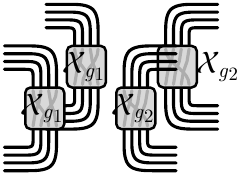}{47}{-20},
\end{equation}
where $\chi_g$ denotes the representation of permutation $g$ acting on the tensor network legs. In this graphical representation, permutations connect replica indices across different copies, making explicit the emergence of an effective classical statistical mechanics model.

The Weingarten function $\text{Wg}_{D}(g)$ can be understood as the inverse of a matrix defined by permutation overlaps. To see this explicitly, consider contracting the unitary gates within each replica on the left-hand side of Eq.~\eqref{eqHaaraverage}. Using unitarity, $U^\dagger U = 1$, this contraction produces the trivial permutation state:
\begin{equation}
\ket{ee}
=
\sum_{g_1,g_2 \in S_{Q}}
\text{Wg}_{D}(g_1^{-1}g_2)
\braket{g_2 g_2 | ee}
\ket{g_1 g_1},
\end{equation}
where $e$ denotes the identity permutation. The overlap between permutation states has a simple combinatorial interpretation. One finds
\begin{equation}
\braket{g g | ee}
=
\braket{g | e}^2
=
D^{C(g)},
\end{equation}
where $C(g)$ denotes the number of cycles in permutation $g$. This result follows from the fact that each cycle contributes an independent index contraction, producing a factor equal to the Hilbert space dimension $D$. Substituting this into the previous expression, we obtain the defining relation for the Weingarten function:
\begin{equation}
\sum_{g_2 \in S_{Q}}
\text{Wg}_{D}(g_1^{-1} g_2)
D^{C(g_2)}
=
\delta_{g_1},
\end{equation}
where $\delta_g = 1$ if $g=e$ and zero otherwise. Together with a similar relation obtained by contracting Eq.~\eqref{eqHaaraverage} with $\bra{ee}$, this shows that the Weingarten functions are the matrix inverse of $D^{C(g)}$ in permutation space. In other words, the Haar average introduces an effective interaction between permutations governed by this inverse structure.

Applying Eq.~\eqref{eqHaaraverage} to the full brick-wall circuit architecture leads to a classical statistical mechanics model defined on the honeycomb lattice. In this mapping, permutation variables $g_i \in S_Q$ live on the vertices of the lattice, and interactions between neighboring vertices arise from contracting unitary gates. Specifically, contracting unitary gates generates an effective Boltzmann weight on links connecting neighboring vertices:
\begin{equation}
W(g_1, g_2)
=
\braket{g_1 | g_2}
=
d^{C(g_1^{-1}g_2)}.
\end{equation}
This expression counts the number of cycles in the relative permutation $g_1^{-1}g_2$. Note that the dimension appearing here is $d$, rather than $D=d^2$, because this weight arises from contractions along a single leg of the unitary tensor. This weight applies to links corresponding to unitary evolution without measurement. Measurements modify this structure: when a measurement occurs, all replicas are projected onto the same local state. After averaging over measurement outcomes, the resulting link weight becomes simply $d$, independent of the permutations.

Since measurements occur with probability $p$, and unitary evolution occurs with probability $1-p$, the effective link weight becomes the weighted average~\cite{Jian2020}
\begin{equation}
W_p(g_1, g_2)
=
(1-p)\, d^{C(g_1^{-1}g_2)}
+
p\, d.
\end{equation}
This expression fully determines the local interaction rules of the effective statistical mechanics model describing monitored Haar random circuits.

Putting everything together, and temporarily ignoring boundary conditions, we obtain the partition function of an anisotropic classical statistical mechanics model defined on the honeycomb lattice:
\begin{equation} \label{eqZHaar}
Z
=
\sum_{\{ g_i \in S_Q \}}
\prod_{ \langle ij \rangle \in G_{s}}
W_p(g_i^{-1} g_j)
\prod_{ \langle ij \rangle \in G_{d}}
\text{Wg}_{D}(g_i^{-1} g_j).
\end{equation}
Here the permutation variables $g_i$ reside on the vertices of the honeycomb lattice. The set $G_s$ denotes solid links, which correspond to contractions along individual tensor legs and incorporate the effects of measurements through the weights $W_p$. The set $G_d$ denotes dashed links, which originate from averaging two-site unitary gates and are weighted by the Weingarten functions. 

As illustrated in Fig.~\ref{fig:statmech}, this mapping transforms the problem of computing entanglement entropies in monitored random circuits into a classical statistical mechanics problem of interacting permutations on the honeycomb lattice. This correspondence provides the foundation for analytical and numerical analysis of measurement-induced entanglement transitions. At this point one might worry that the Boltzmann weights of this statistical mechanics ``model'' can be negative, since they involve Weingarten functions. Regardless of whether this is the case, the replica limit implies that the (conformal) field theory describing the transition is non-unitary: it contains states of negative norm, and the corresponding ``Hamiltonian'' is non-hermitian. This is an intrinsic feature of the problem. Nevertheless, in what follows we will proceed by relying on our intuition from unitary theories.

\subsection{Boundary conditions and domain wall free energy}

\begin{figure}[bt!] 
    \centering
    \includegraphics[width=0.75\textwidth]{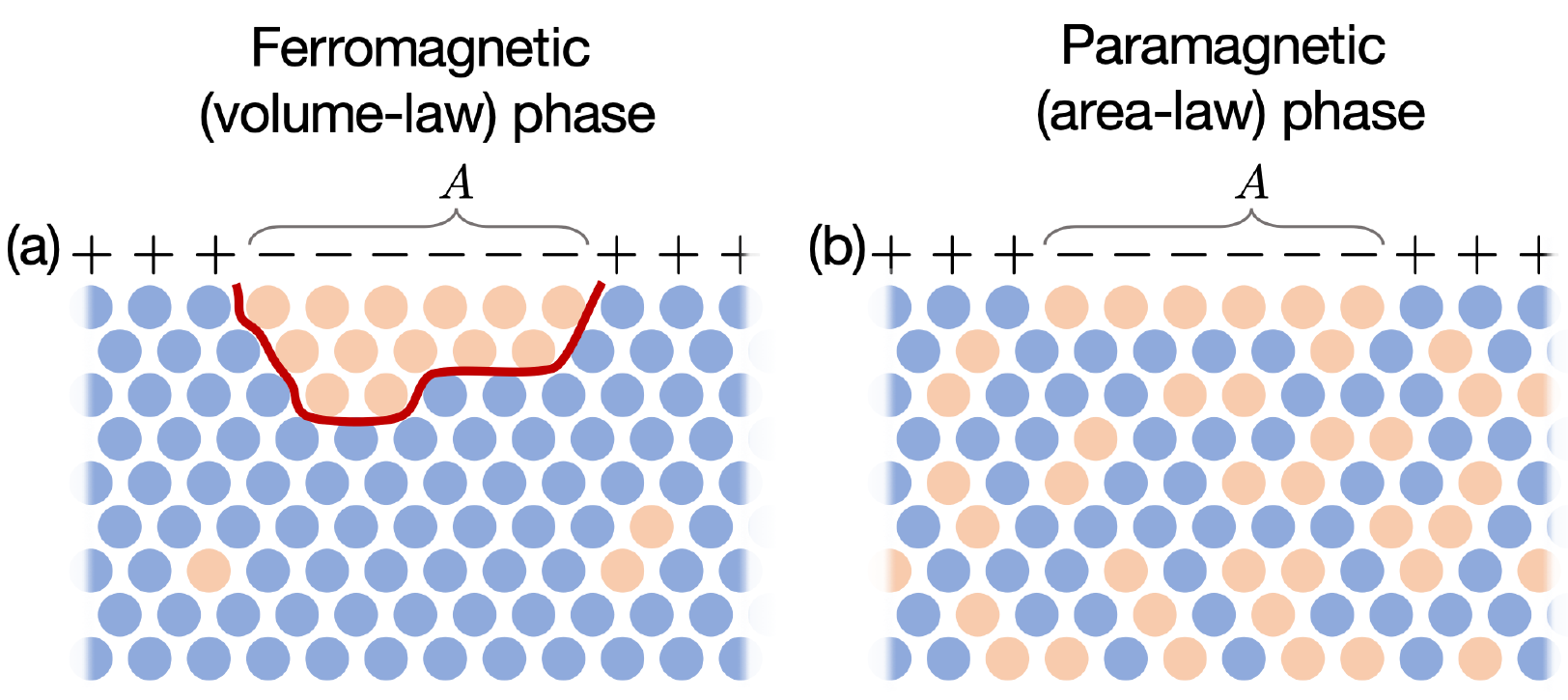}
    \caption{ {\bf Domain wall picture of entanglement dynamics.} Entanglement maps onto the free energy cost of a domain wall at the top boundary in the statistical mechanics model (here shown as an Ising model). At small measurement rate, the statistical mechanics model is ordered, and the free energy cost  $F_A-F_0$ associated with changing the boundary conditions in the region $A$ scales with the size of the interval $L_A$ of $A$ at long times, corresponding to volume-law entanglement. As the measurement rate $p$ is increased, the effective temperature of the statistical mechanics model also increases, leading for $p>p_c$ to a disordered phase with vanishing domain wall tension (corresponding to an area law phase). 
    Adapted from~\cite{Bao2020}.  
    \label{fig:domainwall} 
            }
\end{figure}

By imposing different boundary conditions—corresponding to fixing the permutation variables on the boundary—the statistical mechanics model gives rise to different partition functions. In particular, we define
\begin{equation}\label{eq:Z def}
\begin{split}
Z_A &= \mathbb{E}_{U,\mathbf{m}} \, {\rm tr} \left[{\cal S}_{A,n} \rho_{\mathbf{m}}^{\otimes Q}\right], \\
Z_0 &= \mathbb{E}_{U,\mathbf{m}} \, {\rm tr} \left[\rho_{\mathbf{m}}^{\otimes Q}\right] ,
\end{split}
\end{equation}
where $Z_A$ corresponds to inserting the permutation operator ${\cal S}_{A,n}$ implementing the partial trace over region $A$, while $Z_0$ corresponds to the trivial boundary condition (no permutation insertion). Both partition functions are evaluated in the replicated theory with $Q = nk+1$ copies.

As we showed above, eq.~\eqref{eqEntReplicaTemp}, the averaged $n$th Rényi entropy $\overline{S^{(n)}_A}$ is then obtained in the replica limit via
\begin{equation}
\overline{S^{(n)}_A}
=
\frac{n}{1-n}
\lim_{Q \to 1}
\frac{d}{dQ}
\left(
Z_A - Z_0
\right).
\end{equation}
Here the limit $Q \to 1$ corresponds to $k \to 0$. In this limit one has $Z_A = Z_0 = 1$, since the replicated partition function reduces to a trivial normalization. 

Using this property, the expression for the entropy can be rewritten in a more physically transparent form as a free energy difference:
\begin{equation}
\label{eqFenergycost}
\overline{S^{(n)}_A}
=
\frac{n}{n-1}
\lim_{Q \to 1}
\frac{d}{dQ}
\left(
F_A - F_0
\right),
\end{equation}
where
\begin{equation}
F_A = - \ln Z_A,
\qquad
F_0 = - \ln Z_0 .
\end{equation}
Thus the Rényi entropy is interpreted as the free energy cost associated with changing the boundary condition in region $A$. In the statistical mechanics language, this corresponds to introducing a domain wall at the boundary of $A$. More explicitly, the $S_Q$ ``spins'' of the model are permutation group elements $g_x \in S_Q$ living on the sites of the honeycomb lattice. On the boundary, these spins are pinned by the imposed boundary conditions. For the partition function $Z_0$, the boundary condition is uniform and trivial:
\begin{equation}
g_x = e ,
\end{equation}
for all boundary sites $x$, corresponding to a trivial contraction of replicas. For $Z_A$, the partial trace over subsystem $A$ is implemented by imposing a nontrivial boundary condition on the top boundary:
\begin{equation}\label{eq:SQ boundary}
g_x =
\begin{cases}
g_{\rm SWAP} \equiv (12\cdots n)^{\otimes k}, & x \in A, \\
{\rm identity} = e, & x \in \bar A .
\end{cases}
\end{equation}
This is precisely the same permutation introduced in Eq.~\eqref{eq:boundary def}, but it now appears as a boundary condition in the effective statistical mechanics model. The entropy therefore measures the response of the system to a twist in boundary conditions localized on region $A$.

\vspace{0.5cm}

Having mapped the entanglement entropy calculation in monitored random circuits onto a replica statistical mechanics model, many qualitative aspects of the entanglement transition become transparent (see Fig.~\ref{fig:domainwall}). 

At low measurement rates $p$, the link weights $W_p(g_i^{-1}g_j)$ favor configurations in which neighboring sites carry the same permutation. In other words, the effective interactions are ferromagnetic in permutation space. The statistical mechanics model is then in an \emph{ordered phase}, characterized by long-range alignment of permutation variables. In this phase, introducing the nontrivial boundary condition in region $A$ forces the formation of a domain wall separating regions with different permutation order. This domain wall carries a finite line tension. As a result, the free energy cost
$
F_A - F_0
$
defined in Eq.~\eqref{eqFenergycost} scales extensively with the length $L_A$ of the interval $A$ at long times. Physically, this corresponds to a \emph{volume-law} entanglement phase,
\begin{equation}
\overline{S^{(n)}_A} \sim L_A ,
\end{equation}
in which entanglement grows proportionally to subsystem size.

In contrast, as the measurement rate $p$ approaches 1, measurements dominate over unitary dynamics. In the statistical mechanics description, this effectively increases the temperature and drives the model into a \emph{disordered phase}. In this regime, domain walls proliferate and form a condensate in the bulk. Because bulk domain walls are already abundant, the boundary-imposed domain wall can be absorbed at a finite correlation length from the boundary. Consequently, beyond this length scale, the free energy cost associated with the boundary twist vanishes. The free energy cost of changing the boundary condition therefore scales only with the size of the boundary of $A$ (i.e., it remains finite in one dimension). This leads to \emph{area-law} entanglement scaling:
\begin{equation}
\overline{S^{(n)}_A} \sim \text{const}.
\end{equation}
Thus the measurement-induced entanglement transition corresponds, in the statistical mechanics language, to an order–disorder transition in a model of interacting permutations, with the entanglement entropy playing the role of a boundary free energy.

\subsection{Symmetry and conformal invariance}

A crucial property of the statistical mechanics model derived above (Eq.~\eqref{eqZHaar}) is that the Boltzmann weights are invariant under global left- and right-multiplication of all group elements:
\begin{equation}
\label{LabelEqSymmetry}
g_i \to h_L \, g_i \, h_R^{-1}, \quad 
g_j \to h_L \, g_j \, h_R^{-1}, \qquad h_L,h_R \in S_Q.
\end{equation}
This global symmetry follows from the fact that both the cycle counting function and the Weingarten functions (which are inverses of each other) are {\em class functions}, i.e., they depend only on the conjugacy class of the permutation. Physically, the two factors of $S_Q$ correspond to independent invariance under permuting replicas on the forward-time ($U$) and backward-time ($U^*$) contours. There is also a $\mathbb{Z}_2$ symmetry corresponding to inversion $g \to g^{-1}$. Hence, the global symmetry group of the replicated statistical mechanics model is
\begin{equation}
G = (S_Q \times S_Q) \rtimes \mathbb{Z}_2.
\end{equation}

In the volume-law phase, this left/right permutation symmetry is spontaneously broken down to the diagonal subgroup:
\begin{equation}
S^L_Q \times S^R_Q \ \to \ S_Q,
\end{equation}
corresponding to the transformation $g_i \to h \, g_i \, h^{-1}$ for $h \in S_Q$. In the language of mixed-state transitions, this represents an example of ``strong-to-weak spontaneous symmetry breaking''~\cite{PhysRevX.15.021062,PRXQuantum.6.010344}.

\vspace{0.3cm}

This general mapping indicates that the measurement-induced transition corresponds to a standard ordering, or (replica) symmetry-breaking, transition. Assuming the transition is continuous (second order), it is expected to be described by a two-dimensional Conformal Field Theory (CFT) with central charge $c = 0$ in the replica limit $Q \to 1$. This is because the central charge $c$ measures how the free energy changes under the introduction of a finite scale; since the partition function $Z_0 = 1$ is trivial in the replica limit, we have $c = 0$. Such CFTs with vanishing central charge are non-unitary and notoriously complicated, even in two dimensions. Nevertheless, a number of universal properties follow from scaling and conformal invariance.

\vspace{0.3cm}

Since the bulk properties of the transition depend only on the number of replicas $Q$, this framework naturally explains why all Rényi entropies with $n \ge 1$ exhibit a transition at the same critical measurement rate $p_c$, consistent with numerical observations. Conformal invariance further allows one to extract the general scaling form of the entanglement entropy near criticality. In particular, the ratio of partition functions appearing in Eq.~\eqref{eqFenergycost},
\begin{equation}
\label{LabelEqDefBccOperator}
\frac{Z_A}{Z_0} = \langle \phi_{\rm BCC}(L_A) \, \phi_{\rm BCC}(0) \rangle,
\end{equation}
can be interpreted in CFT language as the two-point function of a boundary-condition-changing (BCC) operator $\phi_{\rm BCC}$ inserted at the endpoints of the entanglement interval $A$~\cite{cardy_conformal_1984,cardy_boundary_2006}. Near criticality, this two-point function scales as
\begin{equation}
\langle \phi_{\rm BCC}(L_A) \, \phi_{\rm BCC}(0) \rangle 
\sim 
\frac{1}{L_A^{2 h(n,k)}} \, f_{n,k} \Big( \frac{L_A}{\xi_Q} \Big),
\end{equation}
where $\xi_Q \sim |p - p_c(Q)|^{-\nu(Q)}$ is the correlation length of the statistical mechanics model, and $f_{n,k}$ is a universal scaling function depending separately on $n$ and $k$. Plugging this expression into the replica formula~\eqref{eqFenergycost}, we obtain the general scaling of the entanglement entropy (up to non-universal additive constants):
\begin{equation}
\overline{S^{(n)}_A} = \alpha_n \ln L_A + f_n \Big( \frac{L_A}{\xi} \Big),
\label{eqEntanglementLog}
\end{equation}
where $\xi \sim |p - p_c|^{-\nu}$ is the correlation length in the $Q \to 1$ limit, and
\begin{equation}
\alpha_n = \frac{2}{n-1} \left. \frac{\partial h}{\partial k} \right|_{k=0},
\end{equation}
is a universal boundary critical exponent. Note that $\alpha_n$ is unrelated to the central charge of the theory; it is instead associated with the scaling dimension of the BCC operator. In particular, conformal invariance predicts that at criticality $p = p_c$, the entanglement exhibits logarithmic scaling,
\begin{equation}
\overline{S^{(n)}_A} \sim \alpha_n \ln L_A ,
\end{equation}
with a universal prefactor that depends on the Rényi index $n$. Importantly, despite the analogy with the scaling of entanglement in the groundstate of 1+1d CFTs, note that the prefactor of the logarithm has nothing to do with the central charge of the underlying CFT: it is actually a boundary scaling dimension, which depends on the R\'enyi index.

\subsection{Large Hilbert Space Dimension Limit}

\subsubsection{Mapping onto Classical Percolation}

In the limit of large on-site Hilbert space dimension, $d \rightarrow \infty$, the $S_Q$ model introduced above reduces to a Potts model with $Q!$ colors defined on the square lattice. To see this, consider the partition function weight $J_p(g_i,g_j;g_k)$ associated with each down triangle in Fig.~\ref{fig:statmech}, obtained by integrating out the middle spin:
\begin{equation}\label{eq:J def}
J_p(g_i,g_j;g_k) = \sum_{g_l \in S_Q} \dia{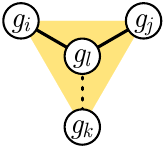}{40}{-20} 
= \sum_{g_l \in S_Q} W_p(g_i^{-1} g_l) W_p(g_j^{-1} g_l) \, \text{Wg}_D(g_l^{-1} g_k),
\end{equation}
as in Eq.~\eqref{eqTriangle} for the two-replica case. The full partition function can then be expressed in terms of these triangle weights:
\begin{equation}\label{eq:Zinfd}
Z = \sum_{\{g_i \in S_Q\}} \prod_{\langle ijk \rangle} J_p(g_i,g_j;g_k),
\end{equation}
subject to the appropriate boundary conditions distinguishing $Z_0$ from $Z_A$.

In the limit $d \to \infty$, we have $d^{C(g)} \sim d^Q \, \delta_g$, where $\delta_g$ is the Kronecker delta on the permutation group $S_Q$: it equals 1 if and only if $g = e$ is the identity permutation, and 0 otherwise. This follows from the fact that the number of cycles $C(g)$ is maximized by the trivial permutation: $C(e) = Q$. Since the Weingarten weights are defined as inverses of $D^{C(g)}$ with $D=d^2$, we immediately obtain
\begin{equation}
\mathsf{Wg}_D(g) \sim D^{-Q} \, \delta_g, \quad d \to \infty.
\end{equation}

Substituting this into the triangle weight~\eqref{eq:J def} and performing straightforward algebra leads to~\cite{Jian2020}:
\begin{equation}\label{eq:J inf d}
J_p(g_i,g_j;g_k) = \big((1-p) \, \delta_{g_i^{-1} g_k} + p \big) \, \big((1-p) \, \delta_{g_j^{-1} g_k} + p \big).
\end{equation}
This factorizes the triangle weight into bond weights on $\langle i k \rangle$ and $\langle j k \rangle$. Explicitly, the weight on bond $\langle i k \rangle$ equals $1$ if $g_i = g_k$ and $p$ if $g_i \neq g_k$, with a similar rule for $\langle j k \rangle$. Interpreting each on-site permutation $g_i \in S_Q$ as a spin state (color), these bond weights match those of a $Q!$-state Potts model on a square lattice, where each unit cell has links between $(i,k)$ and $(j,k)$.

\vspace{0.2cm}

To analytically continue $Q \to 1$, we rewrite the Potts partition function in the Fortuin-Kasteleyn (FK) cluster representation~\cite{FORTUIN1972536}. Each bond weight $(1-p)\delta_{g_i^{-1} g_k} + p$ is interpreted as either an ``occupied'' link corresponding to $(1-p)\delta_{g_i^{-1} g_k}$, or an ``empty'' link corresponding to the trivial term $p$. Occupied links enforce identical permutation spins, forming clusters. Performing the sum over permutations $\sum_{\{g_i \in S_Q\}}$ then gives
\begin{equation}\label{eq:Zpercolation}
Z = \sum_{\rm clusters} p^{\# \text{empty links}} (1-p)^{\# \text{occupied links}} \, \left(Q!\right)^{\# \rm clusters}.
\end{equation}
In this representation, $Q$ only appears in the cluster weight, so taking the replica limit $Q \to 1$ is straightforward. In this limit, all clusters carry trivial weight, and Eq.~\eqref{eq:Zpercolation} reduces to a classical bond percolation problem: links are ``occupied'' with probability $1-p$ (no measurement) and ``empty'' with probability $p$ (corresponding to a local measurement cutting the circuit). This percolation picture naturally captures the measurement-induced transition and predicts critical exponents close to those observed even for small $d$ (e.g., $d=2$ in numerical simulations). For instance, the correlation length diverges as $\xi \sim |p - p_c|^{-4/3}$, with $p_c = 1/2$ in this limit.

\subsubsection{Entanglement and Minimal-Cut Picture}

To study entanglement scaling in this limit, it is convenient to consider a fixed configuration of measurement locations. Averaging over measurement outcomes and Haar gates for a given measurement configuration, the statistical mechanics mapping still applies, with link weights (replacing $W_p$)
\begin{equation}
V_l(g_i^{-1} g_j) = \begin{cases} 
d^{C(g_i^{-1} g_j)}, & \text{if the link is unmeasured},\\[2mm]
d, & \text{if the link coincides with a measurement.} 
\end{cases}
\end{equation}
In the large-$d$ limit,  the unmeasured-link weight satisfies $d^{C(g_i^{-1} g_j)} \sim d^Q \, \delta_{g_i^{-1} g_j}$, so the model becomes a fully ordered (zero-temperature) ferromagnet diluted by measured links: measured bonds are effectively cut, while all unmeasured bonds enforce identical spins. This is consistent with the percolation picture discussed above.

\vspace{0.2cm}

In this limit, a frustrated bond costs a large energy $\sim \ln d$, leading to an effective minimal-cut picture~\cite{Skinner2019}. Recall that computing entanglement involves two partition functions $Z_A$ and $Z_0$, differing only in the boundary conditions at the top of the circuit. The $Z_A$ boundary condition in region $A$ introduces a domain wall (DW) near the top boundary. For $d \to \infty$, the DW follows a minimal cut, i.e., a path cutting the fewest unmeasured links (assumed unique for simplicity). All spins in $Z_0$ are identical, giving $Z_0=1$ trivially.  The DW generates frustrated links, each contributing a large energy:
\begin{equation} 
\Delta E = (n-1) k \ln d,
\end{equation}
using $E_l = -\ln d \, C(g)$ as the link energy. Hence, the DW minimizes the number of unmeasured links it crosses, yielding
\begin{equation}
Z_A = d^{(1-n)k \, \ell_{\rm DW}} \, Z_0,
\end{equation}
where $\ell_{\rm DW}$ is the number of unmeasured links along the minimal cut~\cite{Skinner2019}. In the replica limit, this gives a simple expression for the Rényi entropies:
\begin{equation} 
S_A^{(n)} = \ell_{\rm DW} \, \ln d,
\end{equation}
valid for a fixed measurement configuration. Denoting by $\overline{\ell_{\rm DW}}$ the average over measurement locations (i.e., percolation configurations), we find a simple scaling: $\overline{\ell_{\rm DW}} \sim L_A$ (volume law) for $p < p_{c} = 1/2$, and $\overline{\ell_{\rm DW}} \sim O(1)$ (area law) for $p > p_{c} = 1/2$.

\subsection{Finite $d$ universality class}

The infinite on-site Hilbert space dimension limit discussed above exhibits an {\em accidentally enlarged symmetry}. In this limit, the mapping to the Potts model produces a symmetry group $S_{Q!}$, which is much larger than the generic $(S_Q \times S_Q)$ symmetry of the finite-$d$ Boltzmann weights. (Here, $S_Q \times S_Q$ is a subgroup of $S_{Q!}$: the left- and right-action of $S_Q$ on itself gives a permutation representation $g \in S_{Q!}$, this is due to Cayley's theorem.)  

The leading perturbation that breaks this enlarged symmetry in the Potts model has been identified in Refs.~\cite{PhysRevB.100.134203,Jian2020}. This perturbation is {\em relevant}, with a scaling dimension $\Delta = 5/4$. As a result, for any large but finite on-site Hilbert space dimension $d$, we expect a crossover in the critical behavior. On length scales $\ell \ll \xi(d) \sim d^{4/3}$~\cite{Skinner2019}, the system exhibits percolation-like criticality, reflecting the infinite-$d$ fixed point. On longer length scales $\ell \gg \xi(d)$, the system crosses over to the finite-$d$ universality class, which is independent of $d$ and corresponds to the generic measurement-induced transition.  

This crossover provides a natural explanation for why percolation exponents appear approximately in numerics even for $d = 2$, while the true critical exponents of the measurement-induced transition are only reached at sufficiently large scales.

\section{Discussion}

These lecture notes aimed to provide a concise but self-contained introduction to (replica) statistical mechanics mappings  to random quantum circuits, both in the unitary case and including measurements. The second lecture also contained a short introduction to the physics of measurement-induced phase transitions, adopting a ``learnability'' perspective focusing on the perspective of the observer. In that language, MIPTs are phrased very naturally as phase transitions in terms of how much information the observer can learn from the measurement record. 

Due to time constraints, a number of interesting topics have been left out. Since the original papers, random quantum circuits have become part of the standard toolbox to study chaotic quantum dynamics, and provided crucial insights into, e.g.~thermalization and entanglement growth~\cite{PhysRevX.7.031016,2018arXiv180300089J,PhysRevX.10.031066}, many-body quantum chaos~\cite{PhysRevX.8.041019,PhysRevLett.121.264101,PhysRevLett.121.060601,Friedman2019,PhysRevB.100.064309,PhysRevLett.123.210601} and operator spreading~\cite{Nahum2018,VonKeyserlingk2018}, or the emergence of irreversible hydrodynamics from unitary evolution~\cite{PhysRevX.8.031057,PhysRevX.8.031058}. Meanwhile, the replica statistical mechanics approaches discussed in these notes have also become standard to study random quantum circuits and tensor networks.  Examples of applications include quantum error correction thresholds~\cite{Choi2020,Gullans2019,fan2020self,Li2020b,PhysRevX.11.031066,2021arXiv210804274L}, computational complexity~\cite{2019arXiv190512053H,Napp2019} or to the emergence of unitary ``$k$-designs''~\cite{2019arXiv190512053H}. The field of monitored many-body quantum systems has also become very popular, and some interesting directions not covered in these notes include, to name a few that I find especially interesting (and again, with a non-exhaustive list of references), monitored free fermions~\cite{PhysRevX.13.041045,jian2023measurementinducedentanglementtransitionsquantum}, relation to conformal invariance~\cite{Li2020,2021arXiv210703393Z}, measurement-only dynamics~\cite{Ippoliti2020}, the role of symmetries~\cite{2021arXiv210804274L,2021arXiv210710279A,2021arXiv211109336B}, or very recently the broader relation to learning problems and Bayesian inference~\cite{7dpt-d4s5,295c-lj1w,kim2025measurementinducedphasetransitionsquantum,putz2025learningtransitionsclassicalising}. I refer the interested reader to the review articles~\cite{Potter2022,annurev:/content/journals/10.1146/annurev-conmatphys-031720-030658} for a more detailed account of recent developments.

{\bf Acknowledgements.} I would like to thank all my collaborators and colleagues in this field, from whom I've learned a lot of what I discuss in these notes, including Utkarsh Agrawal, Fergus Barratt, Jacopo De Nardis, Matthew Fisher, Aaron Friedman, Snir Gazit, Sarang Gopalakrishnan, Michael Gullans, Wen Wei Ho, David Huse, Matteo Ippoliti, Chao-Ming Jian, Kabir Khanna, Abhishek Kumar, Yaodong Li, Javier Lopez-Piqueres, Andreas Ludwig, Catherine McCarthy, Adam Nahum, Jed Pixley, Andrew Potter, Hans Singh, Sagar Vijay, Brayden Ware, Justin Wilson, Yi-Zhuang You, and Aidan Zabalo. In particular, I'd like to thank Andrew Potter, with whom I wrote the review~\cite{Potter2022}, which partially overlaps with the discussion of Sec.~\ref{SecStatMech}. I would also like to thank my students Kabir Khanna and Abhishek Kumar for help with the figures of these lectures. I acknowledge support from the Swiss National Science Foundation (grant 10008234) and the Foundation for the University of Geneva (FUNIGE). 

\bibliographystyle{SciPost_bibstyle}
\bibliography{references}

\begin{thebibliography}{10}
\providecommand{\url}[1]{\texttt{#1}}
\providecommand{\urlprefix}{URL }
\expandafter\ifx\csname urlstyle\endcsname\relax
  \providecommand{\doi}[1]{doi:\discretionary{}{}{}#1}\else
  \providecommand{\doi}{doi:\discretionary{}{}{}\begingroup
  \urlstyle{rm}\Url}\fi
\providecommand{\eprint}[2][]{\url{#2}}

\bibitem{Potter2022}
A.~C. Potter and R.~Vasseur,
\newblock \emph{Entanglement Dynamics in Hybrid Quantum Circuits}, pp.
  211--249,
\newblock Springer International Publishing, Cham,
\newblock ISBN 978-3-031-03998-0,
\newblock \doi{10.1007/978-3-031-03998-0_9} (2022).

\bibitem{annurev:/content/journals/10.1146/annurev-conmatphys-031720-030658}
M.~P. Fisher, V.~Khemani, A.~Nahum and S.~Vijay,
\newblock \emph{Random quantum circuits},
\newblock Annual Review of Condensed Matter Physics \textbf{14}(Volume 14,
  2023), 335 (2023),
\newblock \doi{https://doi.org/10.1146/annurev-conmatphys-031720-030658}.

\bibitem{skinner2023lecturenotesintroductionrandom}
B.~Skinner,
\newblock \emph{Lecture notes: Introduction to random unitary circuits and the
  measurement-induced entanglement phase transition} (2023),
  \eprint{2307.02986}.

\bibitem{bertini2026nonequilibriumquantummanybodyphysics}
B.~Bertini,
\newblock \emph{Non-equilibrium quantum many-body physics with quantum
  circuits} (2026), \eprint{2601.22375}.

\bibitem{PhysRevX.7.031016}
A.~Nahum, J.~Ruhman, S.~Vijay and J.~Haah,
\newblock \emph{Quantum entanglement growth under random unitary dynamics},
\newblock Phys. Rev. X \textbf{7}, 031016 (2017),
\newblock \doi{10.1103/PhysRevX.7.031016}.

\bibitem{Nahum2018}
A.~Nahum, S.~Vijay and J.~Haah,
\newblock \emph{{Operator Spreading in Random Unitary Circuits}},
\newblock Physical Review X \textbf{8}(2) (2018),
\newblock \doi{10.1103/PhysRevX.8.021014},
\newblock \eprint{1705.08975}.

\bibitem{PhysRevB.100.134203}
R.~Vasseur, A.~C. Potter, Y.-Z. You and A.~W.~W. Ludwig,
\newblock \emph{Entanglement transitions from holographic random tensor
  networks},
\newblock Phys. Rev. B \textbf{100}, 134203 (2019),
\newblock \doi{10.1103/PhysRevB.100.134203}.

\bibitem{PhysRevB.98.205136}
Y.~Li, X.~Chen and M.~P.~A. Fisher,
\newblock \emph{Quantum zeno effect and the many-body entanglement transition},
\newblock Phys. Rev. B \textbf{98}, 205136 (2018),
\newblock \doi{10.1103/PhysRevB.98.205136}.

\bibitem{Skinner2019}
B.~Skinner, J.~Ruhman and A.~Nahum,
\newblock \emph{{Measurement-Induced Phase Transitions in the Dynamics of
  Entanglement}},
\newblock Physical Review X \textbf{9}(3) (2019),
\newblock \doi{10.1103/PhysRevX.9.031009},
\newblock \eprint{1808.05953}.

\bibitem{Bao2020}
Y.~Bao, S.~Choi and E.~Altman,
\newblock \emph{{Theory of the phase transition in random unitary circuits with
  measurements}},
\newblock Physical Review B \textbf{101}(10) (2020),
\newblock \doi{10.1103/PhysRevB.101.104301},
\newblock \eprint{1908.04305}.

\bibitem{PhysRevLett.129.200602}
F.~Barratt, U.~Agrawal, A.~C. Potter, S.~Gopalakrishnan and R.~Vasseur,
\newblock \emph{Transitions in the learnability of global charges from local
  measurements},
\newblock Phys. Rev. Lett. \textbf{129}, 200602 (2022),
\newblock \doi{10.1103/PhysRevLett.129.200602}.

\bibitem{Jian2020}
C.~M. Jian, Y.~Z. You, R.~Vasseur and A.~W. Ludwig,
\newblock \emph{{Measurement-induced criticality in random quantum circuits}},
\newblock Physical Review B \textbf{101}(10) (2020),
\newblock \doi{10.1103/PhysRevB.101.104302},
\newblock \eprint{1908.08051}.

\bibitem{Zhou2019}
T.~Zhou and A.~Nahum,
\newblock \emph{{Emergent statistical mechanics of entanglement in random
  unitary circuits}},
\newblock Physical Review B \textbf{99}(17) (2019),
\newblock \doi{10.1103/PhysRevB.99.174205},
\newblock \eprint{1804.09737}.

\bibitem{khanna2025randomquantumcircuitstimereversal}
K.~Khanna, A.~Kumar, R.~Vasseur and A.~W.~W. Ludwig,
\newblock \emph{Random quantum circuits with time-reversal symmetry},
\newblock Phys. Rev. Res. \textbf{8}, 013165 (2026),
\newblock \doi{10.1103/4bgj-p2jk}.

\bibitem{RTN}
P.~Hayden, S.~Nezami, X.-L. Qi, N.~Thomas, M.~Walter and Z.~Yang,
\newblock \emph{Holographic duality from random tensor networks},
\newblock Journal of High Energy Physics \textbf{2016}(11), 9 (2016),
\newblock \doi{10.1007/JHEP11(2016)009}.

\bibitem{Preskill2018quantumcomputingin}
J.~Preskill,
\newblock \emph{Quantum {C}omputing in the {NISQ} era and beyond},
\newblock {Quantum} \textbf{2}, 79 (2018),
\newblock \doi{10.22331/q-2018-08-06-79}.

\bibitem{kim2025learningmeasurementinducedphasetransitions}
H.~Kim, A.~Kumar, Y.~Zhou, Y.~Xu, R.~Vasseur and E.-A. Kim,
\newblock \emph{Learning measurement-induced phase transitions using attention}
  (2025), \eprint{2508.15895}.

\bibitem{Gullans2020}
M.~J. Gullans and D.~A. Huse,
\newblock \emph{{Scalable Probes of Measurement-Induced Criticality}},
\newblock Physical Review Letters \textbf{125}(7) (2020),
\newblock \doi{10.1103/PhysRevLett.125.070606},
\newblock \eprint{1910.00020}.

\bibitem{Gullans2019}
M.~J. Gullans and D.~A. Huse,
\newblock \emph{Dynamical purification phase transition induced by quantum
  measurements},
\newblock Phys. Rev. X \textbf{10}, 041020 (2020),
\newblock \doi{10.1103/PhysRevX.10.041020}.

\bibitem{Choi2020}
S.~Choi, Y.~Bao, X.~L. Qi and E.~Altman,
\newblock \emph{{Quantum Error Correction in Scrambling Dynamics and
  Measurement-Induced Phase Transition}},
\newblock Physical Review Letters \textbf{125}(3) (2020),
\newblock \doi{10.1103/PhysRevLett.125.030505},
\newblock \eprint{1903.05124}.

\bibitem{Li2020b}
Y.~Li and M.~P.~A. Fisher,
\newblock \emph{Statistical mechanics of quantum error correcting codes},
\newblock Phys. Rev. B \textbf{103}, 104306 (2021),
\newblock \doi{10.1103/PhysRevB.103.104306}.

\bibitem{fan2020self}
R.~Fan, S.~Vijay, A.~Vishwanath and Y.-Z. You,
\newblock \emph{Self-organized error correction in random unitary circuits with
  measurement},
\newblock Phys. Rev. B \textbf{103}, 174309 (2021),
\newblock \doi{10.1103/PhysRevB.103.174309}.

\bibitem{roberts2017chaos}
D.~A. Roberts and B.~Yoshida,
\newblock \emph{Chaos and complexity by design},
\newblock Journal of High Energy Physics \textbf{2017}(4), 1 (2017).

\bibitem{ORUS2014117}
R.~Or{\'u}s,
\newblock \emph{A practical introduction to tensor networks: Matrix product
  states and projected entangled pair states},
\newblock Annals of Physics \textbf{349}, 117  (2014),
\newblock \doi{https://doi.org/10.1016/j.aop.2014.06.013}.

\bibitem{PhysRevX.15.021062}
R.~Ma, J.-H. Zhang, Z.~Bi, M.~Cheng and C.~Wang,
\newblock \emph{Topological phases with average symmetries: The decohered, the
  disordered, and the intrinsic},
\newblock Phys. Rev. X \textbf{15}, 021062 (2025),
\newblock \doi{10.1103/PhysRevX.15.021062}.

\bibitem{PRXQuantum.6.010344}
L.~A. Lessa, R.~Ma, J.-H. Zhang, Z.~Bi, M.~Cheng and C.~Wang,
\newblock \emph{Strong-to-weak spontaneous symmetry breaking in mixed quantum
  states},
\newblock PRX Quantum \textbf{6}, 010344 (2025),
\newblock \doi{10.1103/PRXQuantum.6.010344}.

\bibitem{cardy_conformal_1984}
J.~L. Cardy,
\newblock \emph{Conformal invariance and surface critical behavior},
\newblock Nuclear Physics B \textbf{240}(4), 514 (1984),
\newblock \doi{10.1016/0550-3213(84)90241-4}.

\bibitem{cardy_boundary_2006}
J.~Cardy,
\newblock \emph{Boundary {Conformal} {Field} {Theory}},
\newblock Encyclopedia of Mathematical Physics  (2006),
\newblock ArXiv: hep-th/0411189.

\bibitem{FORTUIN1972536}
C.~Fortuin and P.~Kasteleyn,
\newblock \emph{On the random-cluster model: I. introduction and relation to
  other models},
\newblock Physica \textbf{57}(4), 536 (1972),
\newblock \doi{https://doi.org/10.1016/0031-8914(72)90045-6}.

\bibitem{2018arXiv180300089J}
C.~{Jonay}, D.~A. {Huse} and A.~{Nahum},
\newblock \emph{{Coarse-grained dynamics of operator and state entanglement}},
\newblock arXiv e-prints arXiv:1803.00089 (2018),
\newblock \eprint{1803.00089}.

\bibitem{PhysRevX.10.031066}
T.~Zhou and A.~Nahum,
\newblock \emph{Entanglement membrane in chaotic many-body systems},
\newblock Phys. Rev. X \textbf{10}, 031066 (2020),
\newblock \doi{10.1103/PhysRevX.10.031066}.

\bibitem{PhysRevX.8.041019}
A.~Chan, A.~De~Luca and J.~T. Chalker,
\newblock \emph{Solution of a minimal model for many-body quantum chaos},
\newblock Phys. Rev. X \textbf{8}, 041019 (2018),
\newblock \doi{10.1103/PhysRevX.8.041019}.

\bibitem{PhysRevLett.121.264101}
B.~Bertini, P.~Kos and T.~Prosen,
\newblock \emph{Exact spectral form factor in a minimal model of many-body
  quantum chaos},
\newblock Phys. Rev. Lett. \textbf{121}, 264101 (2018),
\newblock \doi{10.1103/PhysRevLett.121.264101}.

\bibitem{PhysRevLett.121.060601}
A.~Chan, A.~De~Luca and J.~T. Chalker,
\newblock \emph{Spectral statistics in spatially extended chaotic quantum
  many-body systems},
\newblock Phys. Rev. Lett. \textbf{121}, 060601 (2018),
\newblock \doi{10.1103/PhysRevLett.121.060601}.

\bibitem{Friedman2019}
A.~J. Friedman, A.~Chan, A.~{De Luca} and J.~T. Chalker,
\newblock \emph{{Spectral Statistics and Many-Body Quantum Chaos with Conserved
  Charge}},
\newblock Physical Review Letters \textbf{123}(21), 210603 (2019),
\newblock \doi{10.1103/PhysRevLett.123.210603},
\newblock \eprint{1906.07736}.

\bibitem{PhysRevB.100.064309}
S.~Gopalakrishnan and A.~Lamacraft,
\newblock \emph{Unitary circuits of finite depth and infinite width from
  quantum channels},
\newblock Phys. Rev. B \textbf{100}, 064309 (2019),
\newblock \doi{10.1103/PhysRevB.100.064309}.

\bibitem{PhysRevLett.123.210601}
B.~Bertini, P.~Kos and T.~Prosen,
\newblock \emph{Exact correlation functions for dual-unitary lattice models in
  $1+1$ dimensions},
\newblock Phys. Rev. Lett. \textbf{123}, 210601 (2019),
\newblock \doi{10.1103/PhysRevLett.123.210601}.

\bibitem{VonKeyserlingk2018}
C.~W. {Von Keyserlingk}, T.~Rakovszky, F.~Pollmann and S.~L. Sondhi,
\newblock \emph{{Operator Hydrodynamics, OTOCs, and Entanglement Growth in
  Systems without Conservation Laws}},
\newblock Physical Review X \textbf{8}(2) (2018),
\newblock \doi{10.1103/PhysRevX.8.021013},
\newblock \eprint{1705.08910}.

\bibitem{PhysRevX.8.031057}
V.~Khemani, A.~Vishwanath and D.~A. Huse,
\newblock \emph{Operator spreading and the emergence of dissipative
  hydrodynamics under unitary evolution with conservation laws},
\newblock Phys. Rev. X \textbf{8}, 031057 (2018),
\newblock \doi{10.1103/PhysRevX.8.031057}.

\bibitem{PhysRevX.8.031058}
T.~Rakovszky, F.~Pollmann and C.~W. von Keyserlingk,
\newblock \emph{Diffusive hydrodynamics of out-of-time-ordered correlators with
  charge conservation},
\newblock Phys. Rev. X \textbf{8}, 031058 (2018),
\newblock \doi{10.1103/PhysRevX.8.031058}.

\bibitem{PhysRevX.11.031066}
M.~J. Gullans, S.~Krastanov, D.~A. Huse, L.~Jiang and S.~T. Flammia,
\newblock \emph{Quantum coding with low-depth random circuits},
\newblock Phys. Rev. X \textbf{11}, 031066 (2021),
\newblock \doi{10.1103/PhysRevX.11.031066}.

\bibitem{2021arXiv210804274L}
Y.~Li and M.~P.~A. Fisher,
\newblock \emph{Decodable hybrid dynamics of open quantum systems with
  ${\mathbb{z}}_{2}$ symmetry},
\newblock Phys. Rev. B \textbf{108}, 214302 (2023),
\newblock \doi{10.1103/PhysRevB.108.214302}.

\bibitem{2019arXiv190512053H}
N.~{Hunter-Jones},
\newblock \emph{{Unitary designs from statistical mechanics in random quantum
  circuits}},
\newblock arXiv e-prints arXiv:1905.12053 (2019),
\newblock \eprint{1905.12053}.

\bibitem{Napp2019}
J.~C. Napp, R.~L. La~Placa, A.~M. Dalzell, F.~G. S.~L. Brand\~ao and A.~W.
  Harrow,
\newblock \emph{Efficient classical simulation of random shallow 2d quantum
  circuits},
\newblock Phys. Rev. X \textbf{12}, 021021 (2022),
\newblock \doi{10.1103/PhysRevX.12.021021}.

\bibitem{PhysRevX.13.041045}
M.~Fava, L.~Piroli, T.~Swann, D.~Bernard and A.~Nahum,
\newblock \emph{Nonlinear sigma models for monitored dynamics of free
  fermions},
\newblock Phys. Rev. X \textbf{13}, 041045 (2023),
\newblock \doi{10.1103/PhysRevX.13.041045}.

\bibitem{jian2023measurementinducedentanglementtransitionsquantum}
C.-M. Jian, H.~Shapourian, B.~Bauer and A.~W.~W. Ludwig,
\newblock \emph{Measurement-induced entanglement transitions in quantum
  circuits of non-interacting fermions: Born-rule versus forced measurements}
  (2023), \eprint{2302.09094}.

\bibitem{Li2020}
Y.~Li, X.~Chen, A.~W.~W. Ludwig and M.~P.~A. Fisher,
\newblock \emph{Conformal invariance and quantum nonlocality in critical hybrid
  circuits},
\newblock Phys. Rev. B \textbf{104}, 104305 (2021),
\newblock \doi{10.1103/PhysRevB.104.104305}.

\bibitem{2021arXiv210703393Z}
A.~Zabalo, M.~J. Gullans, J.~H. Wilson, R.~Vasseur, A.~W.~W. Ludwig,
  S.~Gopalakrishnan, D.~A. Huse and J.~H. Pixley,
\newblock \emph{Operator scaling dimensions and multifractality at
  measurement-induced transitions},
\newblock Phys. Rev. Lett. \textbf{128}, 050602 (2022),
\newblock \doi{10.1103/PhysRevLett.128.050602}.

\bibitem{Ippoliti2020}
M.~Ippoliti, M.~J. Gullans, S.~Gopalakrishnan, D.~A. Huse and V.~Khemani,
\newblock \emph{Entanglement phase transitions in measurement-only dynamics},
\newblock Phys. Rev. X \textbf{11}, 011030 (2021),
\newblock \doi{10.1103/PhysRevX.11.011030}.

\bibitem{2021arXiv210710279A}
U.~Agrawal, A.~Zabalo, K.~Chen, J.~H. Wilson, A.~C. Potter, J.~H. Pixley,
  S.~Gopalakrishnan and R.~Vasseur,
\newblock \emph{Entanglement and charge-sharpening transitions in u(1)
  symmetric monitored quantum circuits},
\newblock Phys. Rev. X \textbf{12}, 041002 (2022),
\newblock \doi{10.1103/PhysRevX.12.041002}.

\bibitem{2021arXiv211109336B}
F.~Barratt, U.~Agrawal, S.~Gopalakrishnan, D.~A. Huse, R.~Vasseur and A.~C.
  Potter,
\newblock \emph{Field theory of charge sharpening in symmetric monitored
  quantum circuits},
\newblock Phys. Rev. Lett. \textbf{129}, 120604 (2022),
\newblock \doi{10.1103/PhysRevLett.129.120604}.

\bibitem{7dpt-d4s5}
A.~Nahum and J.~L. Jacobsen,
\newblock \emph{Bayesian critical points in classical lattice models},
\newblock Phys. Rev. B \textbf{112}, 235113 (2025),
\newblock \doi{10.1103/7dpt-d4s5}.

\bibitem{295c-lj1w}
S.~Gopalakrishnan, E.~McCulloch and R.~Vasseur,
\newblock \emph{Monitored fluctuating hydrodynamics},
\newblock Phys. Rev. X \textbf{16}, 011024 (2026),
\newblock \doi{10.1103/295c-lj1w}.

\bibitem{kim2025measurementinducedphasetransitionsquantum}
S.~W. P.~Kim, C.~von Keyserlingk and A.~Lamacraft,
\newblock \emph{Measurement-induced phase transitions in quantum inference
  problems and quantum hidden markov models},
\newblock Phys. Rev. Res. \textbf{8}, 023155 (2026),
\newblock \doi{10.1103/b3s8-y1wb}.

\bibitem{putz2025learningtransitionsclassicalising}
M.~P\"utz, S.~J. Garratt, H.~Nishimori, S.~Trebst and G.-Y. Zhu,
\newblock \emph{Learning transitions in classical ising models and deformed
  toric codes},
\newblock Phys. Rev. Lett. \textbf{136}, 190402 (2026),
\newblock \doi{10.1103/4dwm-kn11}.

\end{thebibliography}
\nolinenumbers

\end{document}